# A Fluid-Structure Interaction Solver for Compressible Flows with Applications in Blast Loading on Thin Elastic Structures

Shantanu Bailoor[1], Aditya Annangi[1], Jung Hee Seo[2], Rajneesh Bhardwaj[1*]

[1]Department of Mechanical Engineering, Indian Institute of Technology Bombay, Mumbai, 400076 India

[2]Department of Mechanical Engineering, Johns Hopkins University, Baltimore, MD, 21218 USA

[*]Corresponding author (email: rajneesh.bhardwaj@iitb.ac.in,

Phone: +91 22 2576 7534, Fax: +91 22 2572 6875)

*Abstract*

We report development and application of a fluid-structure interaction (FSI) solver for compressible flows with large-scale flow-induced deformation of the structure. The FSI solver utilizes partitioned approach to strongly couple a sharp-interface immersed boundary method based flow solver with an open-source finite-element structure dynamics solver. The flow solver is based on a higher-order finite-difference method on Cartesian grid and employs ghost-cell methodology to impose boundary conditions on the immersed boundary. A higher-order accuracy near the immersed boundary is achieved by combining the ghost-cell approach with a weighted least-square error method based on a higher-order approximate polynomial. We present validations for two-dimensional canonical acoustic wave scattering on a rigid cylinder at low Mach number and flow past a circular cylinder at moderate Mach number. The second order spatial accuracy of the flow solver is established by performing a grid refinement study. The structure solver is validated with a canonical elastostatics problem. The FSI solver is validated with published measurements and simulations for the large-scale deformation of a thin elastic steel panel subjected to blast loading in a shock tube. The solver correctly predicts oscillating behavior of the tip of the panel with reasonable fidelity and computed shock wave propagation is qualitatively consistent with the published results. In order to demonstrate the fidelity of the solver and to investigate coupled physics of the shock-structure interaction for a thin elastic plate, we employ the solver for simulating 6.4 kg TNT blast loading on the thin elastic plate. The initial conditions of the blast are taken from field tests reported in the literature. Using numerical schlieren, the shock front propagation, Mach reflection and vortex shedding at the tip of the plate are visualized during the shock wave impact on the plate. We discuss coupling between the non-linear dynamics of the plate and blast loading. The plate oscillates under the influence of blast loading and restoring elastic forces. The time-varying displacement of the tip of plate is found to be superimposition of two dominant frequencies, which correspond to first and second mode of natural frequency of a vibrating plate. The effects of material properties and length of the plate on the flow-induced deformation are briefly discussed. The proposed FSI solver is demonstrated as a versatile computational tool for simulating the blast wave impact on thin elastic structures and given results will be helpful to design thin structures subjected to realistic blast loadings.

**Keywords:** Blast loading, Compressible flows, Fluid-Structure interaction, Immersed boundary method, Flow-induced deformation

# 1 Introduction

Fluid-structure interaction (referred as FSI, hereafter) in compressible flows has applications in several biological as well as engineering systems. Examples of the former include biomechanical interaction of a blast wave with human eyes [1, 2] and brain [3]. One of the active area of research in engineering systems is blast loading on thin structures [4, 5]. The FSI modeling of the deformable structures is challenging since it often involves complex, dynamic boundaries immersed in fluid domain and large-scale flow-induced deformation of the structure. The modeling of the structure domain involves geometric as well as material nonlinearity. The dilatational wave inside the structure plays a critical role in flow-induced structural dynamics and should be numerically resolved for accurate numerical solution. Numerically resolving internal stresses in thin structures may need additional spatial and temporal resolution requirements. The coupling of the governing equations of the fluid and structure may introduce additional non-linearity to the system of equations. The numerical stability of the coupled system of the equations depends on structure-fluid density ratio [6].

Based on the coupling of the flow and structure solvers, the FSI models can be categorized as either utilizing monolithic [7] or partitioned (or segregated) approach [8 - 10]. In the monolithic approach, the governing equations for the flow and structure domains are discretized together, and the non-linear system of equations is solved as a whole. The formulation and numerical solution of such systems become more involved for complex constitutive model for the structure. On the other hand, existing flow and structure solvers can be combined in a partitioned approach. However, a challenge is to implement the data exchange between the two solvers (Fig. 1A). In general, there are two coupling methods commonly used in the partitioned algorithms, namely, explicit (or weak, one-way) coupling and implicit (or strong, two-way) coupling. These methods respectively integrate the governing equations explicitly and implicitly in time. The explicit coupling is computationally inexpensive and may be subject to numerical stability constraints, which depends on the structure-fluid density ratio $\rho_s/\rho_f$ [6]. On the other hand, the implicit coupling is robust, computationally expensive and does not introduce stability constraints. The former is a good choice for larger $\rho_s/\rho_f$ while the latter is needed for smaller $\rho_s/\rho_f$. In the latter case, the structure will respond strongly even to small perturbations from the fluid and vice-versa. Typically, the implicit coupling is needed for the large-scale structure deformation to ensure the numerical stability of the FSI solver. For instance, Bhardwaj and Mittal [8] coupled an in-house incompressible flow solver with an open source structure solver using implicit partitioned approach. Similarly, Tian et al. [9] proposed FSI solver for incompressible flows involving the large-scale flow-induced deformation. In general, better conditioned subsystems result in the partitioned approach as compared to those in the monolithic approach [11]. Heil et al. [12] compared relative performance of the two approaches and concluded that computational time taken in the monolithic approach is on the same order as compared to that in the partitioned approach.

Based on the treatment of the boundary conditions at the fluid-structure interface, the FSI models can be categorized in two divisions: Arbitrary-Lagrangian-Eulerian method (ALE) and immersed boundary method. In the former, Lagrangian formulation of the Navier equation for structure dynamics is solved in a coupled manner with the Eulerian formulation for the Navier-Stokes equations for the flow [7, 13, 14]. The body conformal mesh needs to be mapped by suitable remeshing algorithm at each time step. Tezduyar et al. [15] proposed ALE finite element formulation for the fluid as well as structure domain in which remeshing criterion was based on the mesh size. Similarly, Souli et al. [16] employed ALE formulation with smoothening algorithms to control the mesh quality. However, the large deformation of the fluid-structure interface poses a challenge on the remeshing algorithm. As pointed out by Zheng et al. [6], the remeshing algorithm increases computational time and numerical dissipation is needed to provide robustness in presence of the deformed grid, which could hide effect of under-resolution of grid and degrade solution accuracy. In addition, due to the presence of unstructured mesh, the ALE method does not allow use of geometric multi-grid techniques [6].

Alternatively, in immersed boundary method, initially developed by Peskin [17], the governing equations of the fluid are solved on a fixed Cartesian grid, however, movement of the fluid-structure interface is described in Lagrangian framework. A detailed review of this method was presented by Mittal and Iaccarino [18]. The embedded complex fluid-structure interface in the fluid domain is solved on a non-body conformal Cartesian grid. Previous studies [6, 8, 9, 19, 20] successfully demonstrated the implementation of the immersed boundary method in finite-difference based flow solvers. In general, the interface is treated by diffuse or sharp interface methods. In the former, more recently, De [21] presented diffuse interface immersed boundary method with second order space convergence for solving incompressible flows. In the latter, Ghias et al. [22] presented immersed boundary method based viscous, subsonic compressible flow solver for body non-conformal Cartesian or curvilinear grids with ghost-cell technique to enforce boundary conditions at the immersed boundary. Similarly, Seo and Mittal [23] developed a higher-order, sharp-interface immersed boundary method based solver for low Mach number, flow-induced acoustic waves around complex-shaped rigid bodies. The solver was based on hydrodynamic-acoustic splitting wherein the incompressible flow was first computed using a second-order accurate immersed boundary solver followed by the acoustic component, computed using linearized perturbed compressible equations [24]. Recently, Chaudhuri et al. [25] employed a sharp-interface immersed boundary method based on ghost-cell methodology for two-dimensional shock-obstacle interaction.

Most of the previous reported immersed boundary method based FSI solvers [22, 23, 25] considered rigid, stationary or passive immersed bodies and ignored internal structural stresses. For instance, Eldredge and Pisani [26] considered a passive deformable system in the wake of an obstacle to simulate a fish-like system. To this end, there are two objectives of the present work. First objective is to report the development of a FSI solver for compressible flows involving large-scale flow induced deformation of a thin elastic structure. We build upon our previous works [1, 2] and in the present work, we report several validation cases for the flow solver, structural solver and large-scale flow-induced deformation of a thin elastic plate subjected to the blast loading. Second objective is to investigate coupled physics of shock-structure interaction of a thin elastic plate subjected to realistic blast loading, by employing the solver developed in the first objective. The coupling between non-linear dynamics of the plate and blast loading, and the effect of material properties is investigated in the context of the second objective.

The paper layout is as follows. First, we present the computational modeling of FSI solver in section 2, which couples a sharp-interface immersed boundary method based finite-difference compressible flow solver (section 2.1) with a finite-element structure dynamics solver (section 2.2) using an implicit partitioned approach (section 2.3). Second, we assess the spatial accuracy near the immersed boundary (section 3.1) and present code validation results with the published benchmark data for the flow solver (section 3.2), structural solver (section 3.3) and flow-induced deformation module (section 3.4). The proposed FSI solver is extended to model the non-linear dynamics of a thin elastic plate subjected to an impulsive blast loading by a 6.4 kg TNT charge (section 3.5).

## 2  Computational model

We present the development of a robust, versatile FSI solver which couples compressible, viscous flow solver with an open source finite-element structural solver [27]. The governing equations of the flow domain are solved on a fixed Cartesian (Eulerian) grid while the fluid-structure interface is tracked in Lagrangian framework. The interface is treated using sharp-interface immersed boundary method and the two solvers are coupled using an implicit (two-way) scheme. In the following subsections, we present details of the different modules of the FSI solver.

### 2.1  Fluid dynamics solver

The flow is governed by unsteady, viscous and compressible Navier-Stokes equations. We consider full compressible Navier-Stokes equations, written in conservative form as follows,

$$\frac{\partial \rho}{\partial t} + \frac{\partial (\rho u_i)}{\partial x_i} = 0, \tag{1}$$

$$\frac{\partial (\rho u_i)}{\partial t} + \frac{\partial (\rho u_i u_j)}{\partial x_j} + \frac{\partial p}{\partial x_i} - \frac{\partial \tau_{ij}}{\partial x_j} = 0, \tag{2}$$

$$\frac{\partial e}{\partial t} + \frac{\partial (u_j (e+p))}{\partial x_j} - \frac{\partial (u_k \tau_{jk} + q_j)}{\partial x_j} = 0, \tag{3}$$

$$e = \frac{p}{\gamma - 1} + \frac{1}{2} \rho u_i u_i, \tag{4}$$

where $\rho$, $u_i$, $p$, $\tau_{ij}$, $e$, $q_j$, and $\gamma$ are the density, velocity, pressure, viscous stress tensor, total energy, heat flux and specific heat ratio (1.4 for air), respectively. The dynamic viscosity of the fluid is determined using Sutherland Law. The non-dimensional quantities are normalized with respect to the respective quantities for air at ambient conditions. Eqs. 1-4 are spatially discretized by a sixth-order central compact finite difference scheme [28] and integrated in time using a four-stage Runge-Kutta method. An eighth-order implicit spatial filtering proposed by Gaitonde et al. [29] is applied at the end of each time step to suppress high frequency dispersion errors. In order to resolve the discontinuity in the flow variables caused by a shock wave with the current non-dissipative numerical scheme, the artificial diffusivity method proposed by Kawai and Lele [30] is applied. The viscous stress and heat flux are written as follows,

$$\tau_{ij} = (\mu + \mu^*)\left(\frac{\partial u_i}{\partial x_j} + \frac{\partial u_j}{\partial x_i}\right) + \left(\beta^* - \frac{2}{3}\mu\right)\frac{\partial u_k}{\partial x_k}\delta_{ij}, \quad (5)$$

$$q_j = (\kappa + \kappa^*)\frac{\partial T}{\partial x_j}, \quad (6)$$

$$T = \frac{p}{\rho R}, \quad (7)$$

where $\mu$ and $\kappa$ are physical viscosity and thermal diffusivity, respectively, while $\mu^*$, $\beta^*$ and $\kappa^*$ are artificial shear viscosity, artificial bulk viscosity and artificial thermal diffusivity, respectively. On a non-uniform Cartesian grid, these artificial diffusivities are adaptively and dynamically evaluated as follows,

$$\mu^* = C_\mu \rho \overline{\left|\frac{\partial^4 S}{\partial x_k^4}\right|} \Delta x_k^6, \quad (8)$$

$$\beta^* = C_\beta \rho \overline{\left|\frac{\partial^4 S}{\partial x_k^4}\right|} \Delta x_k^6, \quad (9)$$

$$\kappa^* = C_\kappa \frac{\rho c}{T} \overline{\left|\frac{\partial^4}{\partial x_k^4}\left(\frac{RT}{\gamma-1}\right)\right|} \Delta x_k^5, \quad (10)$$

where $\Delta x_k$ is grid-spacing, overbar denotes Gaussian filtering [31], $C_\mu$, $C_\beta$, and $C_\kappa$ are user-specified constants, $c$ is the speed of sound and $S$ is the magnitude of the strain rate tensor, defined as follows,

$$S_{ij} = \frac{1}{2}\left(\frac{\partial u_i}{\partial x_j} + \frac{\partial u_j}{\partial x_i}\right) \quad (11)$$

We use $C_\mu = 0.002$, $C_\beta = 1.0$ and $C_\kappa = 0.01$, as suggested by Kawai and Lele [30] and fourth derivatives are computed by a fourth-order central compact scheme [28]. It can be noted from Eqs (8)-(10) that the artificial diffusivities are significantly larger only in the region where steep gradient of the flow variables exists, thereby ensuring numerical stability in that region.

The compressible Navier-Stokes equations for the fluid flow with complex structure boundaries inside the fluid domain are solved using the sharp-interface immersed boundary method proposed by Seo and Mittal [23], which was built upon the works of Mittal et al. [19] and Luo et al. [32]. The immersed boundary is represented by an unstructured surface mesh with triangular elements and is embedded in a fluid domain discretized using a non-uniform Cartesian grid. The fluid cells, solid cells and ghost cells are marked with respect to the surface mesh, as described by Mittal et al. [19] and shown in Fig. 2. In the method proposed by Mittal et al. [19], a "normal probe" is extended from the ghost cell to intersect the surface mesh at "body intercept" and any flow quantity near the immersed boundary is computed through a bilinear (trilinear for 3D geometry) interpolation from the surrounding fluid nodes. In the extension proposed by Seo and Mittal [23] and Luo et al. [32] for higher-order accuracy near the immersed boundary, a flow variable $\phi$ near body intercept ($x_{BI}$, $y_{BI}$, $z_{BI}$) is approximated by a $N$th order polynomial $\Phi$ as follows [23, 32],

$$\phi(x',y',z') \approx \Phi(x',y',z') = \sum_{i=0}^{N}\sum_{j=0}^{N}\sum_{k=0}^{N} c_{ijk}(x')^i (y')^j (z')^k, \ i+j+k \leq N \quad (12)$$

where $x' = x - x_{BI}; y' = y - y_{BI}; z' = z - z_{BI}$ and $c_{ijk}$ are unknowns. The number of coefficients required for a $N$th order polynomial are given in Ref. [23]. A third order polynomial is utilized in the present work and the number of coefficients in this case are 10 and 20 for 2D and 3D geometry, respectively. In order to determine the coefficients $c_{ijk}$, we require the values of flow variable $\phi$ at the neighboring fluid points. In order to select these cells, we draw a circle (sphere for 3D cases) of radius $R$ using the body intercept as the center, as discussed in Refs. [23, 32] and illustrated in Fig. 2. If say, $m$ such points are chosen, the coefficients $c_{ijk}$ are determined by minimizing the weighted error $\varepsilon$ defined as follows,

$$\varepsilon = \sum_{n=1}^{N} w_n^2 [\Phi(x',y',z') - \phi(x',y',z')]^2, \quad (13)$$

The weight function $w_n$ is determined by a cosine weight function, suggested by Li [33], as follows,

$$w_n = \frac{1}{2}[1 + \cos(\frac{\pi d_m}{R})] \quad (14)$$

where $d_m$ is the distance between $m^{th}$ point and the body intercept.

## 2.2 Structure dynamics solver

The governing equations for the structure, Navier equations (momentum balance equation in Lagrangian form), are written as follows,

$$\rho_s \frac{\partial^2 d_i}{\partial t^2} = \frac{\partial \sigma_{ij}}{\partial x_j} + \rho_s f_i \quad (15)$$

where $i$ and $j$ range from 1 to 3, $\rho_s$ is the structure density, $d_i$ is the displacement component in the $i$ direction, $t$ is the time, $\sigma_{ij}$ is the Cauchy stress tensor and $f_i$ is the body force component in the $i$ direction. The displacement vector $\mathbf{d}(\mathbf{x}, t)$ describes the motion of each point in the deformed solid as a function of space $\mathbf{x}$ and time $t$. The deformation gradient tensor $F_{ik}$ can be defined in terms of the displacement gradient tensor $\dfrac{\partial d_i}{\partial x_k}$ as follows:

$$F_{ik} = \delta_{ik} + \frac{\partial d_i}{\partial x_k}, \tag{16}$$

where $\delta_{ik}$ is the Kronecker delta, defined as follows,

$$\delta_{ik} = \begin{cases} 1, i = k \\ 0, i \neq k \end{cases} \tag{17}$$

The right Cauchy green tensor is defined in terms of the deformation gradient tensor as follows:

$$C_{ij} = F_{ki}F_{kj} \tag{18}$$

The invariants of the right Cauchy green tensor are defined as follows:

$$\begin{aligned} I_1 &= \lambda_1 + \lambda_2 + \lambda_3 \\ I_2 &= \lambda_1\lambda_2 + \lambda_2\lambda_3 + \lambda_3\lambda_1 \\ I_3 &= \lambda_1\lambda_2\lambda_3 \end{aligned} \tag{19}$$

where $\lambda_i$ are eigenvalues of the right Cauchy green tensor. In the present study, the structure is considered as Saint Venant-Kirchhoff material which considers geometric non-linearity for large-scale deformation for a linear elastic material. For large deformations, the constitutive relation between the stress and the strain is based on Green-Lagrangian strain tensor $\mathbf{E}$ and second Piola-Kirchhoff stress tensor $\mathbf{S}(\mathbf{E})$ as a function of $\mathbf{E}$. The second Piola-Kirchhoff stress tensor can be expressed in terms of Cauchy stress tensor $\sigma$ as follows:

$$\mathbf{S} = J\mathbf{F}^{-1}\sigma\mathbf{F}^{-T} \tag{20}$$

where $J$ is the determinant of the deformation gradient tensor $\mathbf{F}$ and denotes the volume change ratio. The Green-Lagrangian strain tensor $\mathbf{E}$ is defined as,

$$\mathbf{E} = \frac{1}{2}(\mathbf{F}^T\mathbf{F} - 1) \tag{21}$$

The Navier equations are solved by Galerkin finite-element (FE) method for spatial discretization, implemented in Tahoe©, an open-source, Lagrangian, three-dimensional, finite-element solver [27]. It yields the following system of ordinary differential equations for the nodal displacement vector **d**, given by [34],

$$M \ddot{d}_{n+1} + C \dot{d}_{n+1} + K d_{n+1} = F_{n+1} \tag{22}$$

where $M$ is the lumped mass matrix, $C$ is the damping matrix and $K$ is the stiffness matrix. For temporal discretization, Newark method is used, which is a family of integration formulae that depends on two parameters $\beta$ and $\gamma$ [34]:

$$d_{n+1} = d_n + \Delta t \, \dot{d}_n + \frac{\Delta t^2}{2}\{(1-2\beta)\ddot{d}_n + 2\beta \ddot{d}_{n+1}\}$$
$$\dot{d}_{n+1} = \dot{d}_n + \Delta t\{(1-\gamma)\ddot{d}_n + \gamma \ddot{d}_{n+1}\} \tag{23}$$

With $\beta = 0.25$ and $\gamma = 0.5$, an unconditional stable and second order scheme results which used trapezoidal rule. Since Newark method does not account for numerical damping, undesired or spurious high-frequency oscillations are not handled effectively by this method [34, 35]. In the present paper, we employ Hilber-Hughes-Taylor (HHT) time integration scheme (or $\alpha$-method) developed by Hilber et al. [36] and this method accounts for the numerical damping with a parameter $\alpha$ in the governing equation such that [34],

$$M \ddot{d}_{n+1} + (1+\alpha)C\dot{d}_{n+1} - \alpha C \dot{d}_n + d\sigma + (1+\alpha)K d_{n+1} - \alpha K d_n = F(t_{n+\alpha}) \tag{24}$$

$$t_{n+\alpha} = (1+\alpha)t_{n+1} = t_{n+1} + \alpha \Delta t \tag{25}$$

If $\alpha$, $\beta$ and $\gamma$ are selected such that $\alpha \in [-1/3, 0]$, $\gamma = (1 - 2\alpha)/2$ and $\beta = (1 - \alpha)^2/4$, an unconditionally stable, second order accurate scheme results [34]. Decreasing $\alpha$ increases the amount of numerical damping and $\alpha = 0$ corresponds to the Newark method. In the present simulations, we use $\alpha = -0.30$, $\beta = 0.4225$ and $\gamma = 0.80$, allowing for the maximum numerical damping in the finite element model.

## 2.3 Fluid-structure interaction coupling

The compressible flow solver (section 2.1) and structural solver (section 2.2) are coupled using the implicit partitioned approach. Note that an in-house incompressible flow solver [19] was coupled with the structural dynamics solver Tahoe[©] [27] by Bhardwaj and Mittal [8] and we implement the implicit coupling reported in Ref. [8], in the present work. The solvers are coupled such that they exchange data at each time step (Fig. 1A). The flow solution is marched by one time step with the current deformed shape of the structure and the velocity of the fluid-structure interface act as the boundary condition in the flow solver (Fig. 1B). The boundary condition representing the continuity of velocity at the interface (or no slip on the structure surface) is as follows,

$$u_{i,f} = \dot{d}_{i,s} \qquad (26)$$

where subscripts $f$ and $s$ denote the fluid and structure, respectively. The pressure loading on the structure surface exposed to the fluid domain is calculated using the interpolated normal fluid pressure at the boundary intercept points via a trilinear interpolation (bilinear interpolation for 2D), as described by Mittal et al. [19]. This boundary condition represents continuity of the traction at the solid-fluid interface and is expressed as follows,

$$\sigma_{ij,f} n_j = \sigma_{ij,s} n_j \qquad (27)$$

where $n_j$ is the local surface normal pointing outward from the surface. The structural solver is marched by one-time step with the updated fluid dynamic forces. The convergence is declared after the $L_2$ norm of the displacement of the fluid-structure interface reduces below a preset value [8]. In order to ensure the numerical stability of the FSI solver at low structure-fluid density ratio, under-relaxation of the displacement and the velocity of the fluid-structure interface is implemented, as discussed by Bhardwaj and Mittal [8].

## 3 Results and Discussion

In order to test and validate the proposed FSI solver, we perform several tests, described in the following sections. First, we perform a grid refinement study to establish the second order accuracy near the immersed boundary; second, we present independent validations of the flow solver, structural solver and flow-induced deformation module. Finally, we employ the solver to simulate the blast loading on a thin elastic plate.

### 3.1 Test for spatial accuracy of the FSI solver

The spatial accuracy of the immersed boundary method was tested by simulating a low speed, subsonic flow past a circular cylinder. The Mach number of the free-stream and Reynolds number are 0.2 and 45, respectively. At such low Reynolds numbers, viscous forces are strong enough to prevent flow separation downstream of the cylinder, leading to a steady wake (Fig. 3 (a)). This state can be effectively used to compute the spatial accuracy of the solver. Similar tests were carried out by Mittal et al. [19] and Ghias et al. [22]. The problem was simulated over several uniform Cartesian grids of dimension ($8d \times 8d$), where $d$ is the cylinder diameter, and resolutions ranging from 0.1 to 0.01 (Fig. 3 (b)). Since an analytical solution does not exist for this problem, the result in each case is compared against one obtained over a highly refined grid with grid size 0.005. The variation of $L_2$ norms of density and velocity errors with grid spacing is plotted in Fig. 4. The scaling of error follows a second-order trend, demonstrating the design order of accuracy of the immersed boundary method.

### 3.2 Validation of the flow solver

#### 3.2.1 Acoustic scattering at low Mach numbers

We test the flow solver against published results of acoustic wave scattering from a circular cylinder [37, 38]. A cylinder of diameter 1 is placed with its center at (0,0) and subjected to a Gaussian pressure pulse at (4,0). The initial intensity of the perturbation pulse is given by:

$$p'(t=0) = \in \exp\left(-\ln(2)\frac{(x-4)^2 + y^2}{0.2^2}\right) \tag{28}$$

In the above equation, $\in$ is a constant with value $10^{-3}$. We employed a $600 \times 600$ uniform Cartesian grid over a domain of [-6, 6] with $\Delta x = \Delta y = 0.02$ at a time step of 0.01. No slip-wall and Neumann boundary condition are applied on the cylinder surface for velocity and pressure, respectively. Non-reflecting energy transfer and annihilation boundary condition, developed by Edgar and Visbal [39], are applied at the flow domain boundaries. Fig. 5 shows wave propagation in the fluid medium and acoustic pulse scattering/reflection from the cylinder surface at different time intervals. The left column plots the data of the present work while right column shows data of Liu and Vasilyev [38]. In Fig. 5A, a principal pulse is generated due to the initial pressure perturbation. The acoustic wave strikes the cylinder and a part of the wave reflects off the cylinder surface leading to a secondary acoustic wave (Fig. 5B). The principal wave front continues propagating towards the far boundary of the domain. As explained by Liu and Vasilyev [38], the two parts of the principal wave front, split by the cylinder, traverse its span, collide and merge, that results in a third acoustic wave front generation (Fig. 5C). The third wave then propagates towards the far boundaries of the domain. Qualitatively, the similar features of the wave scattering and reflection of the acoustic waves were reported by Liu and Vasilyev [38] (right column in Fig. 5). We place five numerical probes around the cylinder, at coordinates (2,0), (2,2), (0,2), (-2,2) and (-2,0), as shown in Fig. 6A, and record the perturbation pressure at each probe. The pressures are plotted and are compared with data of Liu and Vasilyev [38] in Fig. 6B-F. The comparisons are in excellent agreement and deviations in the pressure perturbations are within 5% of the values reported by Liu and Vasilyev [38].

### 3.2.2 Flow past a circular cylinder

In order to benchmark the compressible flow solver, we compute the flow past a circular cylinder at several Reynolds numbers ($Re$) and Mach number of 0.2. Fig. 7 plots vorticity contours at different $Re$ which shows typical von Kármán vortex street in all cases. Fig. 8 plots temporal variation of the pressure drag and lift coefficients, defined respectively as follows, $C_D = 2F_D/(\rho_\infty U_\infty^2 D)$ and $C_L = 2F_L/(\rho_\infty U_\infty^2 D)$, where $F_D$ and $F_L$ are the pressure drag and lift force, respectively, $D$ is the diameter of the cylinder, $\rho_\infty$ is the density of air at ambient and $U_\infty$ is the free stream velocity. The time-varying behavior of $C_D$ and $C_L$ is attributed to the von-Karman vortex shedding and is qualitatively consistent with the previous data. We compute pressure coefficient on the cylinder surface, defined as, $C_p = 2(p_{mean} - p_\infty)/(\rho_\infty U_\infty^2)$, where $p_{mean}$ is the time-averaged pressure computed after the vortex shedding reaches a dynamic-stationary state and $p_\infty$ is the atmospheric pressure. Fig. 9A compares $C_p$ with data of Ghias et al. [22] at $Re = 300$ and $Re = 1000$ with respect to azimuthal angle, $\theta$ (defined in the inset of Fig. 9A). In addition, in Fig. 9B, we compare base pressure coefficient $C_{pb}$, defined as $C_p$ at lee side of the cylinder at point P (shown in the inset of Fig. 9A) with data of Ghias et al. [22], Henderson [40] and Williamson and Roshko [41], at several $Re$. The computed values of $C_p$ as well as $C_{pb}$ are in good agreement with those reported in the previous studies [22, 40, 41]. As explained by Ghias et al. [22], the discrepancy with the experimental data of Williamson et al. [41] at $Re > 180$ in Fig. 9B is attributed to intrinsically three-dimensional flow fields, which is not captured by our two-dimensional simulation. We also note a good match in the comparison of computed Strouhal number ($St = fD/U_\infty$, where $f$ is the frequency of time-varying lift coefficient) at several $Re$ in Fig. 10A with data of Ghias et al. [22]. Similarly, the computed time-averaged values of the pressure drag coefficient at several $Re$ in Fig. 10B are in good agreement with previous studies [22, 40]. Overall, the comparisons are good and validate the flow solver.

### 3.3 Validation of structural solver

We validate the structural solver against benchmark problem in which an infinitely long annulus is subjected to a radial displacement of *s* at inner surface and zero traction at the outer surface [32, 42]. The exact solution of the benchmark can be obtained for static, linearly elastic problem using axisymmetric plane-strain Lamé equation, given by [32, 42],

$$\sigma_{\theta\theta} = \frac{A}{r^2} + 2C, \qquad (29)$$

with the boundary conditions defined as follows,

$$d(r) = s \text{ at } r = R_1,$$
$$\sigma_{\theta\theta} = 0 \text{ at } r = R_2, \qquad (30)$$

where the inner and outer radius of the annulus are $R_1$ and $R_2$, respectively. The exact solution for radial displacement *d(r)* of Eq. 30 is given by [32, 42],

$$d(r) = -\frac{A}{2G}\frac{1}{r} + \frac{2\nu C}{\lambda}r, \qquad (31)$$

where Lamé coefficients, $\lambda$ and *G*, are defined in terms of Young's modulus, *E*, and Poisson's ratio, *ν*, as follows,

$$\lambda = \frac{\nu E}{(1+\nu)(1-2\nu)}, \quad G = \frac{E}{2(1+\nu)} \qquad (32)$$

The constants *A* and *C* are defined as,

$$A = \frac{2GR_1^2 R_2}{R_2^2 - R_1^2}, \quad C = \frac{A}{2R_1^2(1-2\nu)} \qquad (33)$$

The contours of the radial displacement *d(r)* calculated using the structure solver are plotted in Fig. 11A for the following parameters, $R_1 = 1$, $R_2 = 2$, $E = 20$, $\nu = 0.33$, and $s = 0.05$. Fig. 11B compares the computed numerical solution and exact solution obtained using Eq. 31. The comparison is excellent and validates the structure solver implemented in the FSI solver.

### 3.4 Validation of flow-induced deformation module for blast loading

In our previous study [2], we validated computed pressure loadings (incident as well as reflected) obtained from the flow solver against free field TNT blast measurements. In addition, validation of the pressure loadings for a low energy blast was carried out by Bhardwaj et al. [1] for small-scale deformation of the structure. In this section, we validate the FSI solver for the large-scale deformation of the structure subjected to blast loading against measurement reported by Giordano et al. [43]. In this measurement, a shock wave of Mach number 1.21 impacts on a deformable 50 mm thin steel panel of thickness 1 mm in quiescent air, as shown in schematic in Fig 12 (a). The plate is mounted on 15 mm high non-deformable base and authors used shadowgraph technique to visualize the shock front propagation and panel deformation [43]. In this context, previous numerical studies [44-48] also considered this measurement [43] for the validation of FSI solver.

In order to model the measurement of Giordano et al. [43], we utilize a computational domain, considered in Ref. [47] and is shown in Fig. 12 (a). The left part of the shock tube is initialized such that a shock wave of Mach number of 1.21 is produced in quiescent air (101 kPa, 293 K). The initial pressure, density and velocity on the left part in Fig. 12 (a) are obtained using Rankine-Hugoniot relations and calculations of the initial values are given in Appendix. We employ a non-uniform Cartesian grid with stretching, as shown in Fig 12 (b). The grid is uniform ($\Delta x = 0.085$ mm, $\Delta y = 0.2$ mm) in the region in which the panel is expected to move and non-uniform grid stretching is used from this region to the downstream. A zoomed-in view of the grid in the vicinity of the immersed boundary is shown in the inset of Fig. 12 (b). The panel is discretized using quadrilateral finite-elements, as shown in the inset of Fig. 12 (b). The length and thickness of the plate are discretized using 400 and 10 elements. The time-step for the flow as well as structural solver is $7.3 \times 10^{-9}$ s. The finite-deformation Saint-Venant Kirchhoff material model described in section 2.2 is utilized and material properties of the steel panel used in the model are as follows: density = 7600 kg/m$^3$, Young's Modulus = 220 GPa and Poisson's ratio = 0.3.

The interaction of the blast wave with the panel is quantified by plotting the time-varying displacement of the tip of the plate. The computed time-varying tip displacement is plotted in Fig 13 along with the data in previously reported numerical and experimental published studies [43-45] . Simulation results are shown by solid or broken lines while the measurement is shown by filled squares. The tip displacement signal shows that the plate moves forward due to high pressure loading generated by the impact of the blast wave and later move backwards due to elastic restoring forces, representing the motion of the tip by a sinusoidal-like curve. Our simulation shows similar oscillating behavior of the tip displacement, as reported in numerical studies by Giordano et al. [43], Sanches and Coda [44], Deiterding and co-workers [45-47] and Li et al. [48]. The difference between the computed forward and backward displacement at different time instances with respect to previously reported simulations [43, 44] is on the order of 10-23%. The rate of deformation of the panel is consistent as compared to that in earlier simulations [43, 44], during the forward as well as backward motion in Fig. 13. The present computed signal shows a slight lag of around 0.4 ms in timings of maximum forward and backward displacement, as compared to earlier simulations [43, 44].

The dominant frequency in the displacement signal is obtained using FFT analysis and the time-period based on this frequency is 2.6 ms. The time-period in published simulations [43, 44] is around 2.8 ms, that is very close to the computed value in the present work. The natural frequency ($f_{ni}$) of the vibration of a cantilever beam obtained using Euler–Bernoulli beam theory is as follows [49, 50],

$$f_{ni} = \frac{k_i^2}{2\pi}\sqrt{\frac{EI}{\rho_s A L^4}} \qquad (34)$$

where $EI$, $\rho_s$, $A$ and $L$ are flexural rigidity, density, cross-sectional area and length of the beam respectively. $i = 1, 2, 3$ represents frequency modes and $k$ is the respective constant for the modes. The values of $k$ are 1.875, 4.694 and 7.855 for first, second and third mode of the natural frequency, respectively. Using Eq. 34, the calculated time periods of the oscillation ($1/f_{ni}$) in the first and second mode are 2.9 and 0.5 ms, respectively. Since the simulated time period is closer to the first mode of the natural frequency, the plate oscillates in the first mode.

As compared to numerical data of Deiterding and Wood [45], the differences in the maximum forward displacement and oscillation frequency in the present work as well as in Refs. [43, 44] are on the order of 30-40% and 10-15%, respectively. This may be explained by the fact that Deiterding and Wood [45] used a thin-shell finite element solver, while our study as well as others studies [43, 44] employed a linear elastic material with consideration of geometric non-linearity.

The rate of deformation of the panel matches well with the measurement [43] during the forward motion (0 < $t$ < 1 ms) in Fig. 13. The error in the simulated maximum forward displacement with respect to the published experiment [43] is around 15% (Fig. 13). There is a discrepancy between the present as well as earlier simulations [43, 44] and the measurement after panel starts turning backwards ($t$ > 1.8 ms). Giordano et al. [43] attributed it to movement of the base on which the panel is mounted, during the shock propagation, which results in additional displacement of the panel.

Further, we qualitatively compare the simulated shock wave propagation with corresponding numerical and experimental results of Giordano et al. [43] at similar time instances. In Fig. 14 (first column), we plot contours of the magnitude of the density gradient i.e. numerical schlieren, during initial impact of the shock wave on the panel. The respective numerical and experimental results of Giordano et al. [43] are plotted in second and third column of Fig 14, respectively. After the shock wave impinges on the panel ($t = 0$ μs), the transmitted and reflected shock waves appear in front and behind the panel, respectively, along with Mach reflection from the corner of the base. The Mach reflection of the shock wave by the base and vortex shedding from the tip of the panel due to roll off of separating shear layer are noted at after $t = 127$ μs (Fig. 14, first column). The simulated wave propagation qualitatively agrees with published numerical and experimental results. In particular, the shapes of the reflected and transmitted waves, along with the formation of the vortex at the tip of the panel are in agreement with published results.

Overall, our FSI solver correctly predicts the oscillating behavior of the plate and the error for the maximum forward displacement with respect to the measurement is around 15%. The time-varying tip displacement signal and shock wave propagation are also consistent with the published simulations. The latter is also in qualitative agreement with published experiments. Therefore, the flow-induced deformation module of the FSI solver for thin structures involving large-scale deformation is validated with reasonable fidelity.

### 3.5  Realistic blast loading on a thin elastic plate

In order to demonstrate the capability and application of the FSI solver developed, we consider 6.4 kg TNT blast at on thin elastic aluminum plate of length 0.235 m mounted at a distance of 3.175 m from the center of the blast (Fig. 15 (a)). The plate is fixed at the bottom and the ratio of the plate thickness to its length is 0.02. Fig. 15 (a) illustrates the computational domain with initial conditions and non-reflecting boundary condition [39] is applied at all the boundaries except at the bottom boundary where slip condition is imposed. Bailoor et al. [2] obtained initial conditions of the blast using free field measurements of the explosion of 6.4 kg TNT given in Bentz and Grimm [51]. We employ the same initial conditions in the simulations presented in this section and are plotted in Fig. 15 (b). The finite-deformation Saint Venant Kirchhoff model described in section 2.2 is utilized as the material model. The material properties used for the aluminum plate are as follows: density = 2700 kg/m$^3$, Young's Modulus = 70 GPa and Poisson's ratio = 0.35. Since $\rho_s/\rho_f$

~ O(1000) for metal plates, we employ explicit coupling for the simulations presented in this section. The grid sizes near the plate are $\Delta x = \Delta y = 0.94$ mm and the length and thickness of the plate are discretized using 400 and 10 quadrilateral elements. The time-step for the flow as well as structural solver is $3.4 \times 10^{-8}$ s.

Fig. 16 and 17 plot the pressure contours (left column) and numerical schlieren (right column) for the impact and interaction of the blast wave with the plate. The numerical schlieren are plotted using the contours of the magnitude of the density in the computational domain. The blast wave propagates in air (0 ms) and impacts on the plate (0.4 ms), as plotted in Fig. 16. The plate acts a bluff body to the air flow with stagnation point at the base of the plate. A large pressure develops on the plate in the upstream (left column, 0.4 ms) near the stagnation point. The shock wave gets transmitted in the downstream and reflects off the plate (0.4 to 0.7 ms, right column in Fig. 16). A vortex starts forming due to roll-up of shear layers on the lee side of the plate (0.5 to 0.7 ms, right column). A low pressure region near the plate tip seen in left column from 0.5 to 0.7 ms corroborates the formation of the vortex. In Fig. 17, the shock wave further travels to downstream and Mach reflection with bottom boundary can be noted (1 ms, right column). Vortices are also generated from the base of the plate in the upstream (1 ms to 1.7 ms, left column) and the tip vortex grows bigger in size while the transmitted shock travels to downstream, as seen in right column from 0.9 ms to 1.7 ms in Fig. 17. At 1.7 ms, a significant bending of the plate is seen with sub-atmospheric pressure in the upstream region near the plate. The simulated pressure field in left column of Figs. 16 and 17 shows qualitatively similar features to with the one obtained by Shi et al. [52] for the blast loading on rigid concrete column, verifying the numerical results qualitatively.

Fig. 18 plots vorticity contours around the plate at time instances $t_1 = 1$ ms, $t_2 = 1.7$ ms and $t_3 = 3.1$ ms and the vortex shedding after the blast wave impingement at 1.7 ms on the plate tip can be noted, as discussed earlier. Vortex shedding at the tip of the plate was also visualized in the experiments reported by Giordano et al. [43], verifying the numerical results qualitatively. The vortices are also generated from the base of the plate in the upstream at 1.7 ms, as discussed earlier.

The flow-induced deformation of the plate due to blast loading is quantified by plotting the time-varying tip of the plate (top left corner of the plate, shown as red dot on the plate in Fig. 15 (a)). Fig. 19 (a) shows oscillating behavior of the tip displacement of the plate in horizontal direction - the tip moves forward and further backwards with respect to its initial position due the elasticity of the plate. The FFT analysis of the signal shows two dominant frequencies and time periods corresponding to these frequencies are 18.1 ms and 2.3 ms. The time periods obtained from the modal analysis (Eq. 34) are 14.3 ms and 2.3 ms in the first and second mode of the natural oscillation, respectively. Since the time period in first as well as second mode obtained by the modal analysis is respectively very close to those obtained by the FFT analysis, the signal of the plate displacement is superposition of the first and second mode of the natural oscillation. The first time period corresponds to overall displacement signal of the plate, similar to a sinusoidal-like curve and it is superimposed by secondary oscillations due to second time period. In this section, we consider plate length, $L = 235$ mm, and the plate of smaller length ($L = 50$ mm) showed only first mode of oscillation in section 3.4. Therefore, the plate length is an important parameter in deciding the modes of the oscillation, consistent with findings for an oscillating elastic splitter plate in Ref. [50].

In Fig 19 (b), we further quantify the response of the plate under blast loading by plotting the contours of the principal stress in the plate, $s_1$, at different time instances (shown as black squares in Fig. 19 (a)). The inset of Fig. 19 (b) at $t_1$ shows the finite-element mesh inside the plate. In the simulation, we resolve propagation of dilatational wave across the plate thickness and the time-varying stresses in Fig. 19 (b) confirm the wave propagation inside the plate. At instance of the maximum forward displacement ($t_4$), a large stress appears in the middle left region of the plate due to larger bending of the plate.

Finally, we tested the FSI solver for a steel plate with the following material properties: density = 7600 kg/m$^3$, Young's Modulus = 220 GPa and Poisson's ratio = 0.30. Fig. 20 compares tip displacement for aluminum and steel plates, in which steel plate exhibits lesser deformation that that of aluminum due to its larger Young's modulus. The magnitude of the maximum and minimum displacement is around 65% lesser in the former as compared to that in the latter. The FFT analysis show that the signal of the tip displacement of the steel plate also shows two dominant frequencies and thereby, the plate oscillates with the superimposition of the first and second mode of the natural oscillation, as discussed earlier for the aluminum plate.

## 4    Conclusions

We describe a fluid-structure interaction (FSI) solver for compressible flows to investigate blast loading on thin elastic structures. The numerical model employs implicit partitioned approach to couple a sharp-interface immersed boundary method based compressible flow solver and an open source finite-element structure solver. The flow solver exhibits second order spatial accuracy and is validated against two benchmarks - acoustic wave scattering past a rigid cylinder at low Mach number and flow past a circular cylinder at moderate Mach number. The validation of the structural solver is presented for a canonical elastostatics problem. The flow-induced deformation module of the solver is validated against published numerical simulations and measurement of the blast loading on a thin steel panel mounted in a shock tube. The validations are carried out for time-varying displacement of the tip of the plate and qualitative visualization of shock front during the impact of the shock wave on the panel. Overall, the computed quantities in all test cases are in good agreement with previous numerical and experimental results and validate the present FSI solver. The FSI solver is employed to investigate the impact of a blast wave on a thin aluminum and steel plates for 6.4 kg TNT charge at 3.175 m from the plate. The length and thickness of the plates are 0.235 m and 4.7 mm, respectively. We compute time-varying flow field, pressure field, flow-induced structural deformation and internal plate stresses during the blast loading. The propagation of the shock front, Mach reflection and vortex shedding at tip of the plate are visualized using numerical schlieren. The displacement of the tip of the plate is oscillatory –the tip of the plate undergoes a forward displacement due to the impact of the blast wave and after reaching the maximum displacement, it moves backwards due to restoring elastic forces. The FFT analysis show two dominant frequencies, which correspond to the first and second mode of the natural oscillation of a cantilever beam. The first mode determines overall displacement curve of the plate, similar to a sinusoidal-like curve and it is superimposed by secondary oscillations due to the second mode of the oscillation. The plate deformation is influenced by the material properties of the plate, and magnitude of the maximum and minimum displacement of the steel plate is around 65% lesser as compared to that of the aluminum plate. Overall, the FSI solver is able to capture the blast wave propagation, its non-linear interaction with the plate and associated structure dynamics with reasonable fidelity. The present results and insights will also be useful to design thin elastic structures subjected to realistic blast loadings.

# 5 Acknowledgements

R.B. gratefully acknowledges financial support from Department of Science and Technology (DST), New Delhi through fast track scheme for young scientists. This work was also partially supported by an internal grant from Industrial Research and Consultancy Centre (IRCC), IIT Bombay. We thank Professor Ralf Deiterding at University of Southampton for answering several email queries and an anonymous reviewer for useful comments. R.B. thanks Professor B. Puranik at IIT Bombay and Prof. Thao D. Nguyen at Johns Hopkins University for useful discussions.## 6 Appendix

In this appendix, we show calculations of the pressure ($P$), velocity ($U$) and temperature ($T$) behind the shock using Rankine-Hugoniot relations. The pressure behind the shock ($P_2$) is given by [53],

$$\frac{P_2}{P_1} = 1 + \frac{2\gamma}{\gamma+1}(M_1^2 - 1) \qquad (35)$$

where subscript 1 and 2 correspond to conditions behind and ahead of the shock, respectively, $\gamma$ is ratio of specific heats of the air and $M_1$ is Mach number of the shock wave. Substituting $P_1 = 101$ kPa, $\gamma = 1.4$, $M_1 = 1.21$ in Eq. 35, we obtain $P_2 = 155.7$ kPa. The density behind the shock ($\rho_2$) is expressed as follows [53],

$$\frac{\rho_2}{\rho_1} = \frac{(\gamma+1)M_1^2}{2+(\gamma-1)M_1^2} \qquad (36)$$

Substituting $\rho_1 = 1.2$ kg/m$^3$, $\gamma = 1.4$ and $M_1 = 1.21$ in Eq. 36, we obtain, $\rho_2 = 1.63$ kg/m$^3$. The Mach number behind the shock is given by [53],

$$M_2 = \sqrt{\frac{1 + \frac{(\gamma-1)}{2}M_1^2}{\gamma M_1^2 - \frac{(\gamma-1)}{2}}} \qquad (37)$$

Substituting $\gamma = 1.4$ and $M_1 = 1.21$ in Eq. 37, we get $M_2 = 0.836$. The temperature behind the shock is found by utilizing the equation of state, $p = \rho RT$ and is given as follows,

$$\frac{T_2}{T_1} = \frac{P_2}{P_1}\frac{\rho_1}{\rho_2} \qquad (38)$$

Substituting the values of the pressure and density behind the shock (obtained using Eq. 35 and 36, respectively) as well as ahead of the shock, and $T_1 = 293$ K in Eq. 38, we get, $T_2 = 332.5$ K. The speed of sound behind the shock is expressed as follows,

$$a_2 = \sqrt{\gamma R T_2} \qquad (39)$$

where $R$ is gas constant of air, $R = 287$ J·kg$^{-1}$·K$^{-1}$. Substituting values of $\gamma$, $R$ and $T_2$ in Eq. 39, we get $a_2 = 365.5$ m/s. Therefore, the velocity behind the shock is, $U_2 = M_2 a_2 = 305.6$ m/s. Similarly, the velocity ahead of the shock is expressed as,

$$U_1 = M_1 a_1 = M_1 \sqrt{\gamma R T_1} \qquad (40)$$

After substituting values of $M_1$, $\gamma$, $R$ and $T_1$ in eq. 40, we get, $U_1 = 415.2$ m/s. Since the Rankine-Hugoniot equations are based on a frame of reference moving with the shock, the velocity behind the shock in a fixed frame of reference in the simulation is, $U_{2,\text{rel}} = U_1 - U_2 = 109.6$ m/s. The pressure and density behind the shock are 155.7 kPa and 1.63 kg/m$^3$, obtained from eq. 35 and 36, respectively.

# 7 References

1. Bhardwaj R, Zeigler K, JH Seo, and KT Ramesh, Nguyen TD (2014) A Computational Model of Blast Loading on Human Eye, *Biomechanics and Modeling in Mechanobiology*, Vol. 13 (1), pp 123-140,.
2. Bailoor S, Bhardwaj R, Nguyen TD (2015) Effectiveness of Eye armour during blast loading, *Biomechanics and Modeling in Mechanobiology*, Vol. 14(6), pp 1227-1237,
3. Ganpule S, Alai A, Plougonven E, Chandra N (2013) Mechanics of blast loading on the head models in the study of traumatic brain injury using experimental and computational approaches, *Biomechanics and Modeling in Mechanobiology*, Vol. 12 (3), pp 511 - 531.
4. Kambouchev, N., L. Noels, R. Radovitzky (2006) Nonlinear compressibility effects in fluid-structure interaction and their implications on the air-blast loading of structures, *Journal of Applied Physics*, Vol. 100, pp 063519 (1 -11),
5. Wang E, Gardner N, Shukla A (2009) The blast resistance of sandwich composites with stepwise graded cores, *International Journal of Solids and Structures,* Vol. 46 (18), pp 3492-3502.
6. Zheng X., Xue Q., Mittal R. and Beilamowicz S. (2010). A Coupled Sharp-Interface Immersed Boundary-Finite-Element Method for Flow-Structure Interaction with Application to Human Phonation", *Journal of Biomechanical Engineering*. Vol. 132, 111003 (1-12).
7. Dunne T. and Rannacher R. (2006). Adaptive Finite Element Approximation of fluid-structure interaction based on an Eulerian Variational Formulation. *Lecture Notes in Computational Science and Engineering* (Eds: Bungartz H.-J. and Schaefer M.), Springer Verlag.
8. Bhardwaj R, Mittal R (2012) Benchmarking a Coupled Immersed-Boundary-Finite-Element Solver for Large-Scale Flow-Induced Deformation. *AIAA Journal* 50 (7):1638-1642.
9. Tian, F.-B., et al., (2014) Fluid–structure interaction involving large deformations: 3D simulations and applications to biological systems. *Journal of Computational Physics*, Vol. 258: p. 451-469.
10. Schafer M., Heck M. and Yigit S. (2006). An implicit Partitioned Method for the Numerical Simulation of Fluid-Structure Interaction. *Lecture Notes in Computational Science and Engineering* (Eds: Bungartz H.-J. and Schaefer M.), Springer Verlag.

11. Förster C, Wall W, Ramm E (2007) Artificial mass instabilities in sequential staggered coupling of nonlinear structures and incompressible viscous flows. *Comput Methods Appl Mech Eng* 196:1278–1293.
12. Heil M., Hazel A. L. and Boyle J. (2008). Solvers for large-displacement fluid–structure interaction problems: segregated versus monolithic approaches. *Comput Mech*. Vol. 43, 91-101.
13. Hron J and Turek S. (2006). A Monolithic FEM/Multigrid Solver for an ALE formulation of Fluid-Structure Interaction with Applications in Biomechanics. *Lecture Notes in Computational Science and Engineering* (Eds: Bungartz H.-J. and Schaefer M.), Springer Verlag.
14. Sahin M. and Mohseni K. (2009). An arbitrary Lagrangian–Eulerian formulation for the numerical simulation of flow patterns generated by the hydromedusa Aequorea victoria. *J. Comp. Phys*. 228. 4588.
15. Tezduyar TE, Sathe S, Stein K and Aureli L. (2006). Modeling of fluid-structure interactions with the space-time techniques. *Lecture Notes in Computational Science and Engineering* (Eds: Bungartz H.-J. and Schaefer M.), Springer Verlag.
16. Souli M, Ouahsine A and Lewin L. (2000). ALE formulation for fluid-structure interation problems, *Compt Methods Appl. Mech. Engrg*. Vol 190, pp 659-675.
17. Peskin, C. S. (1977) Numerical analysis of blood flow in the heart, *Journal of Computational Physics*, Vol. 25, pp. 220-252.
18. Mittal, R., Iaccarino, G. Immersed Boundary Methods (2005) *Annual Review of Fluid Mechanics*, , Vol 37, pp 239-61, 2005.
19. Mittal R, Dong H, Bozkurttas M, Najjar FM, Vargas A, vonLoebbeck A. (2008) A Versatile Immersed Boundary Methods for Incompressible Flows with Complex Boundaries. *J Comp Phys* 227 (10).
20. Mittal, R., X. Zheng, R. Bhardwaj, J.H. Seo, Q. Xue, and S. Bielamowicz (2011) Toward a simulation-based tool for the treatment of vocal fold paralysis. *Frontiers in Computational Physiology and* Medicine, Vol. 2.
21. A. K. De (2016) A diffuse interface immersed boundary method for convective heat and fluid flow. International Journal of Heat and Mass Transfer Vol. 92, 957–969.
22. Ghias, R., Mittal, R., Dong, H. (2007) A Sharp Interface Immersed Boundary Method for Compressible Viscous Flows. *Journal of Computational Physics*. Vol. 225, pp. 528-553.
23. Seo J. H. and Mittal R. (2011) A high-order immersed boundary method for acoustic wave scattering and low-Mach number flow-induced sound in complex geometries. *J Comp Phys* Vol. 230, 1000–1019.
24. Seo, J. H. and Moon, Y. J. (2006) Linearized Perturbed Compressible Equations for Low Mach Number Aeroacoustics," *Journal of Computational Physics*, Vol. 218, pp. 702-719.
25. Chaudhuri A, Hadjadj A, Chinnayya A (2011) On the use of immersed boundary methods for shock/obstacle interactions. *J Comp Phys* Vol. 230, pp 1731–1748.
26. Eldredge J D and Pisani D (2008) Passive locomotion of a simple articulated fish-like system in the wake of an obstacle, J. Fluid Mech. vol. 607, pp. 279–288.
27. Tahoe is an open source C++ finite element solver, which was developed at Sandia National Labs, CA (http://sourceforge.net/projects/tahoe/).
28. Lele S (1992) Compact finite Difference Schemes with Spectral-like Resolution. *J Comp Phys* Vol. 103, pp 16-42, 1992
29. Gaitonde D, Shang JS, Young JL. (1999) Practical Aspects of Higher-order Accurate Finite Volume Schemes for Wave Propagation Phenomena. *International Journal for Numerical Methods in Engineering*, Vol. 45:1849-1869,
30. Kawai S and Lele SK (2008) Localized Artificial Diffusivity Scheme for Discontinuity Capturing on Curvilinear meshes. *J Comp Phys* Vol. 227, pp 22, 2008

31. Cook AW and Cabot WH (2005) Hyper Viscosity for Shock-turbulence Interactions. *J Comp Phys* 203(2):379-385.
32. H. Luo, R. Mittal, X. Zheng, S. A. Bielamowicz, R. J. Walsh, J. K. Hahn (2008) An immersed-boundary method for flow–structure interaction in biological systems with application to phonation. *Journal of computational physics* 227.22: 9303-9332.
33. Li, Z. (1998) A fast iterative algorithm for elliptic interface problems." *SIAM Journal on Numerical Analysis* 35.1 : 230-254.
34. Hughes T. J. R. (1987). The Finite Element method. Prentice-Hall. Englewood Cliffs, NJ.
35. Negrut D., Rampalli R., Ottarsson G. and Sajdak A. (2007). On an Implementation of the Hilber-Hughes-Taylor Method in the Context of Index 3 Differential-Algebraic Equations of Multibody Dynamics (DETC2005-85096). Journal of Computational and Nonlinear Dynamics. Vol. 2, 73.
36. Hilber H. M., T.J.R. Hughes, and R.L. Taylor, (1977) Improved numerical dissipation for time integration algorithms in structural dynamics, Earthquake Engineering and Structural Dynamics 5, 283–292.
37. C.K.W. Tam, J.C. Hardin (Eds.), (1997) Second Computational Aeroacoustics (CAA) Workshop on Benchmark Problems, NASA-CP-3352.
38. Liu Q and Vasilyev OV (2007) A Brinkman penalization method for compressible flows in complex geometries. *J Comp Phys* 227 (2007) 946–966.
39. Edgar NB and Visbal MR (2003) A General Buffer Zone Type Non-reflecting Boundary Condition for Computational Aeroacoustics. *AIAA Paper* 2003-3300, 2003.
40. Henderson, R.D., (1995) Details of the drag curve near the onset of vortex shedding. Physics of Fluids, 7(9): p. 2102-2104.
41. Williamson C.H.K., Roshko A., (1990) Measurement of base pressure in the wake of a cylinder at low Renolds numbers. Zeitschrift fuer Flugwissenschaften und Weltraumforschung, 14: p. 38-46.
42. Fung, Y.C., (1965) Foundation of Solid Mechanics, Prentice-Hall, Englewood Cliffs, NJ.
43. Giordano, J., G. Jourdan, Y. Burtschell, M. Medale, D. E. Zeitoun, L. Houas (2005) Shock wave impacts on deforming panel, an application of fluid-structure interaction." *Shock Waves* 14.1-2 (2005): 103-110.
44. Sanches RAK, Coda HB (2014) On fluid-shell coupling using an arbitrary Lagrangian-Eulerian fluid solver coupled to a positional Lagrangian shell solver, *Applied Mathematical Modelling*, Vol. 38, pp 3401-3418.
45. Deiterding R., S. Wood (2013) Parallel adaptive fluid-structure interaction simulation of explosions impacting on building structures, Computers & Fluids, 88: 719–729.
46. Deiterding R., (2007) An adaptive level set method for shock-driven fluid-structure interaction Proc. Appl. Math. Mech. 7(1): 2100037–2100038. doi:10.1002/pamm.200700258
47. Deiterding, R., F. Cirak, S. P. Mauch. (2009) Efficient Fluid-Structure Interaction Simulation of Viscoplastic and Fracturing Thin-Shells Subjected to Underwater Shock Loading." *International Workshop on Fluid-Structure Interaction. Theory, Numerics and Applications*. kassel university press GmbH, pp 65 - 80.
48. Li Q, Liu P, He G (2015) Fluid-solid coupled simulation of the ignition transient of solid rocket motor, *Acta Astronautica*, Vol. 110, pp 180-190.
49. Thomson W. T., Dahleh M. D. (1997) Theory of vibration with application, 5$^{th}$ edition. Prentice Hall, Upper Saddle River, New Jersey.
50. Kundu A. K., Soti A. K. Bhardwaj R., Thompson M. (2017) The Response of an Elastic Splitter Plate Attached to a Cylinder to Laminar Pulsatile Flow, Journal of Fluids and Structures, Vol. 68, Pages 423–443.
51. Bentz V, Grimm G (2013) Joint live fire (JLF) final report for assessment of ocular pressure as a result of blast for protected and unprotected eyes (Report number JLF-TR-13-01) U.S. Army Aberdeen Test Center, Aberdeen Proving Ground, MD.

52. Shi, Y., Hao H. and Li Z.-X. (2007) Numerical simulation of blast wave interaction with structure columns. *Shock Waves* Vol. 17, pp 113-133.
53. Anderson J. D. (1990) Modern Compressible Flow, Second edition, Mc-Graw Hill, New York.

# 8 Figures

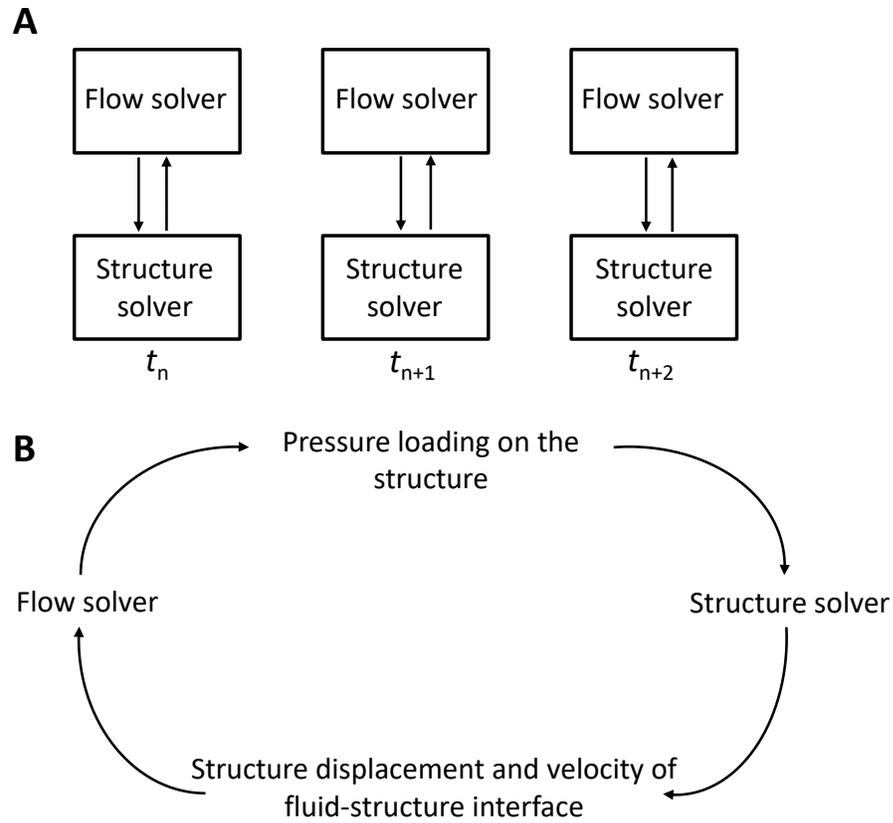

Fig. 1: (A) Partitioned approach showing data exchange between the flow and structural solver utilized in the present work (B) Data exchange between the solvers at fluid-structure interface.

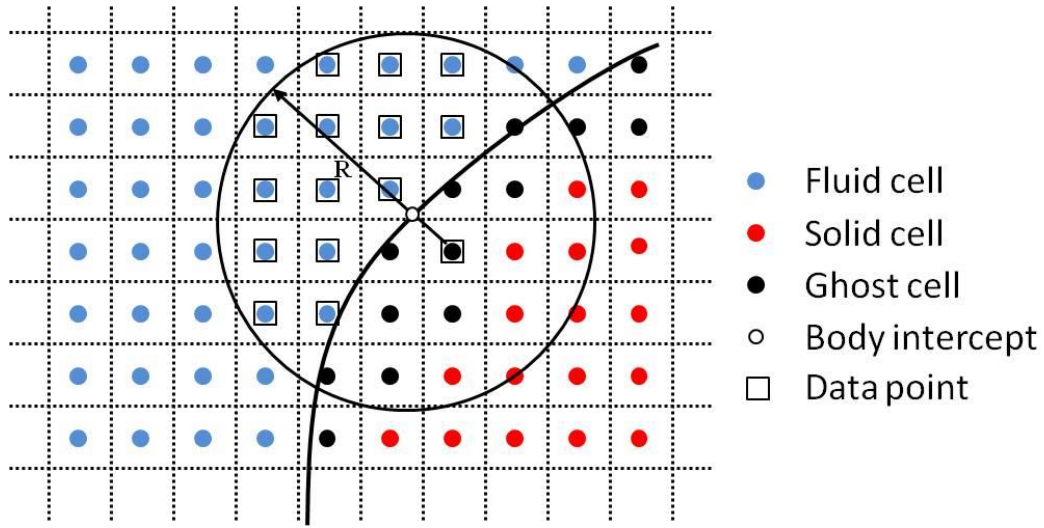

Fig. 2: Schematic of the ghost-cell method proposed by Seo and Mittal [23] and used in the present compressible flow solver.

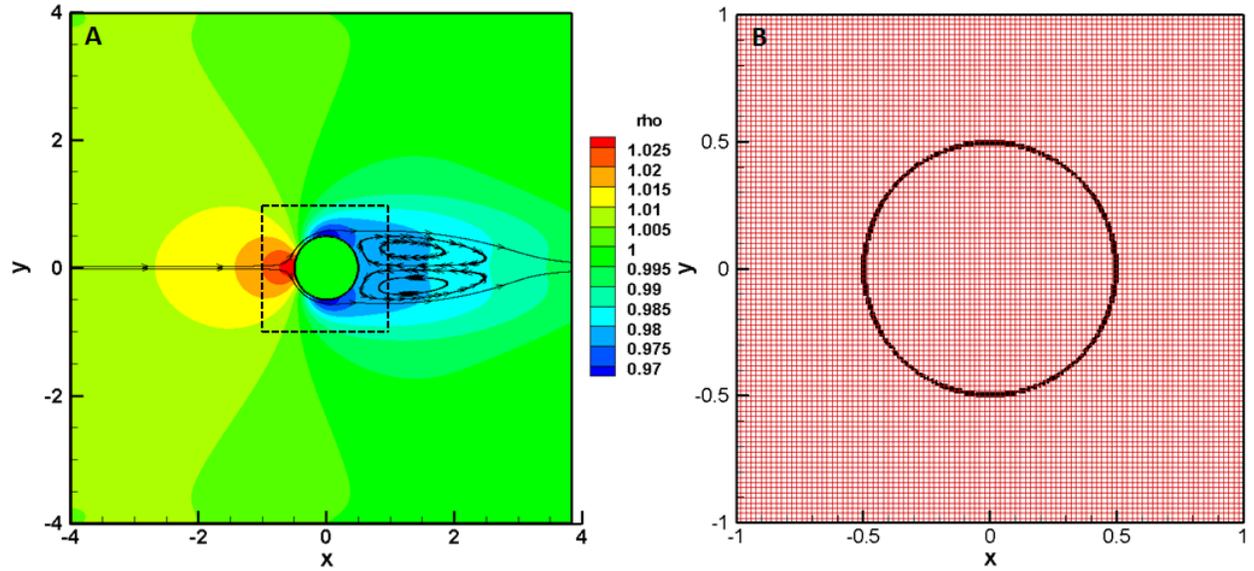

Fig. 3: (a) Layout of spatial accuracy test indicating the steady wake in downstream of the cylinder and (b) Outline of cylinder mapped over one of the tested grids with $\Delta x = \Delta y = 0.02$.

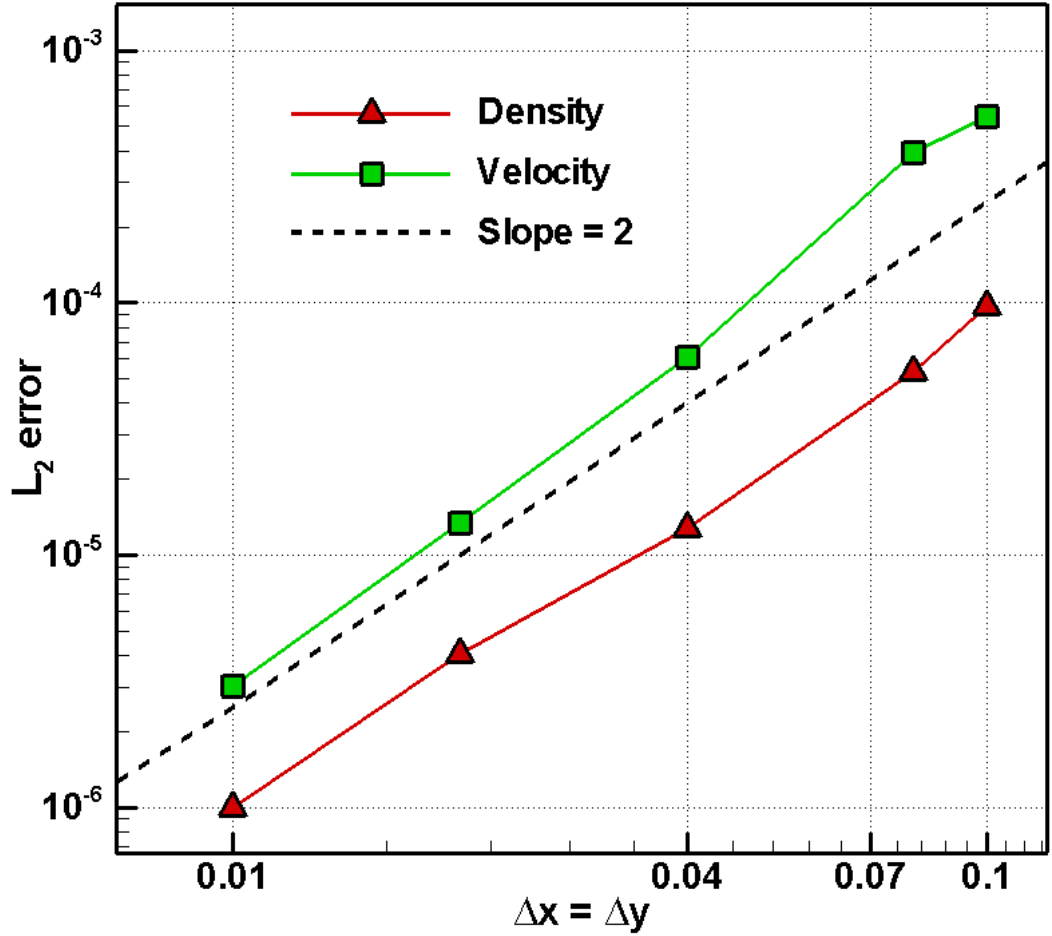

Fig. 4: Plots of $L_2$ error norm of density and velocity against grid spacing for the spatial accuracy test. A dotted line of slope = 2 is plotted for reference.

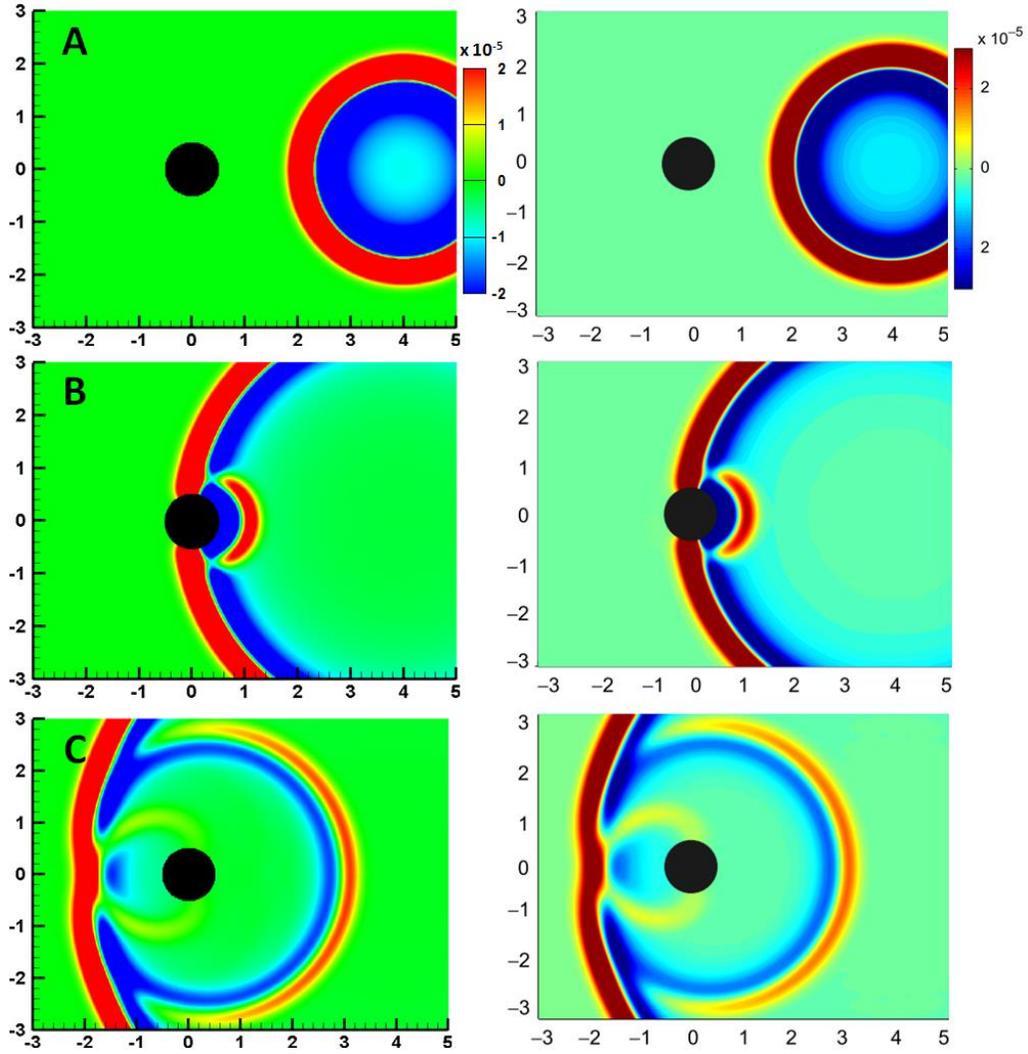

Fig. 5: Qualitative comparison of pressure perturbation contours between the results of the present study (left column) and Liu and Vasilyev [38] (right column) at times (A) $t = 2.0$, (B) $t = 4.0$ and (C) $t = 6.0$. Images in right column reprinted with permission from [38]. Copyright (2007) Elsevier.

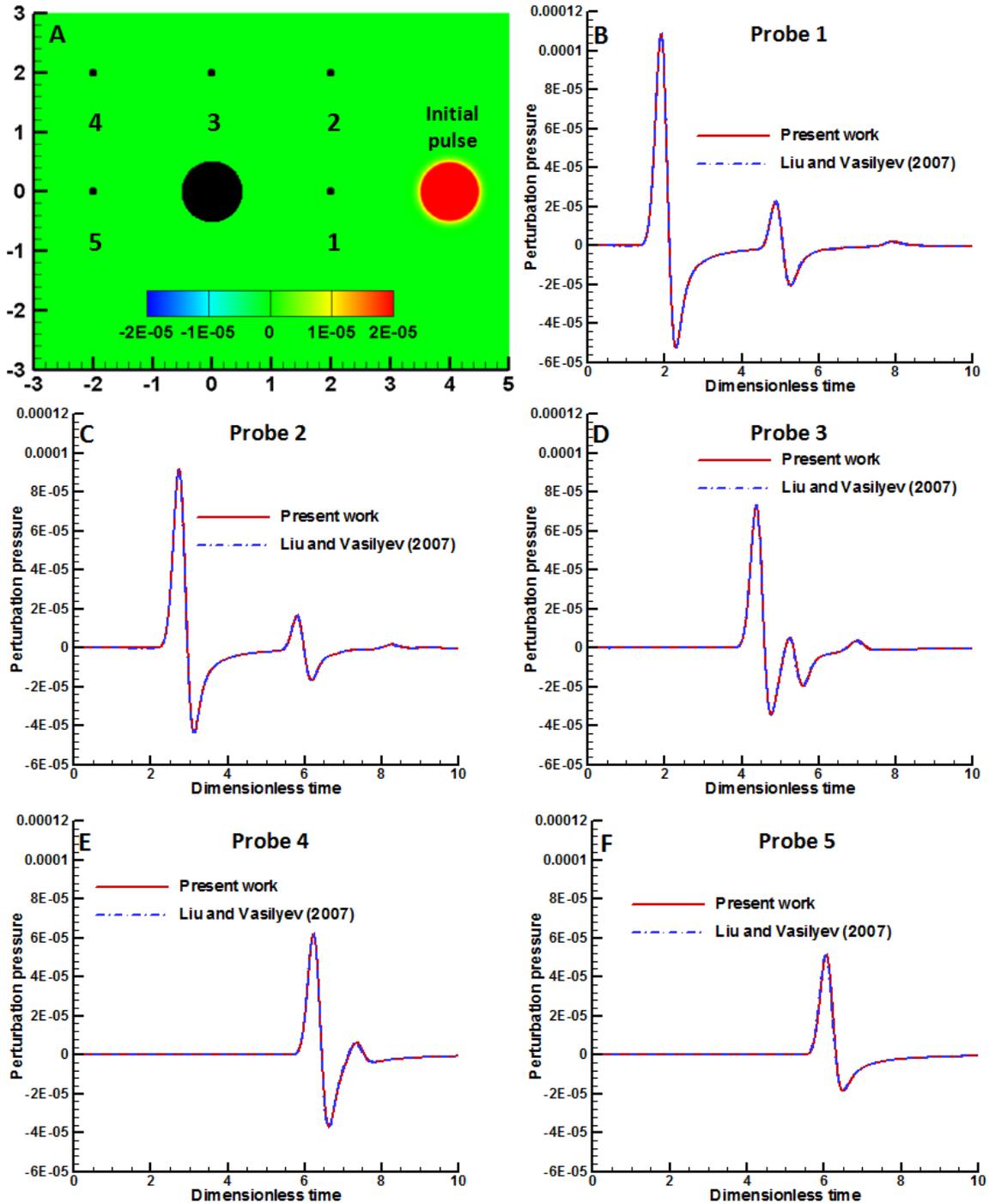

Fig. 6: (A) Locations of the probes with respect to the cylinder and initial acoustic pulse (B-F) Comparison between time-varying pressure recorded at the five probes obtained in the present study and reported by Liu and Vasilyev [38].

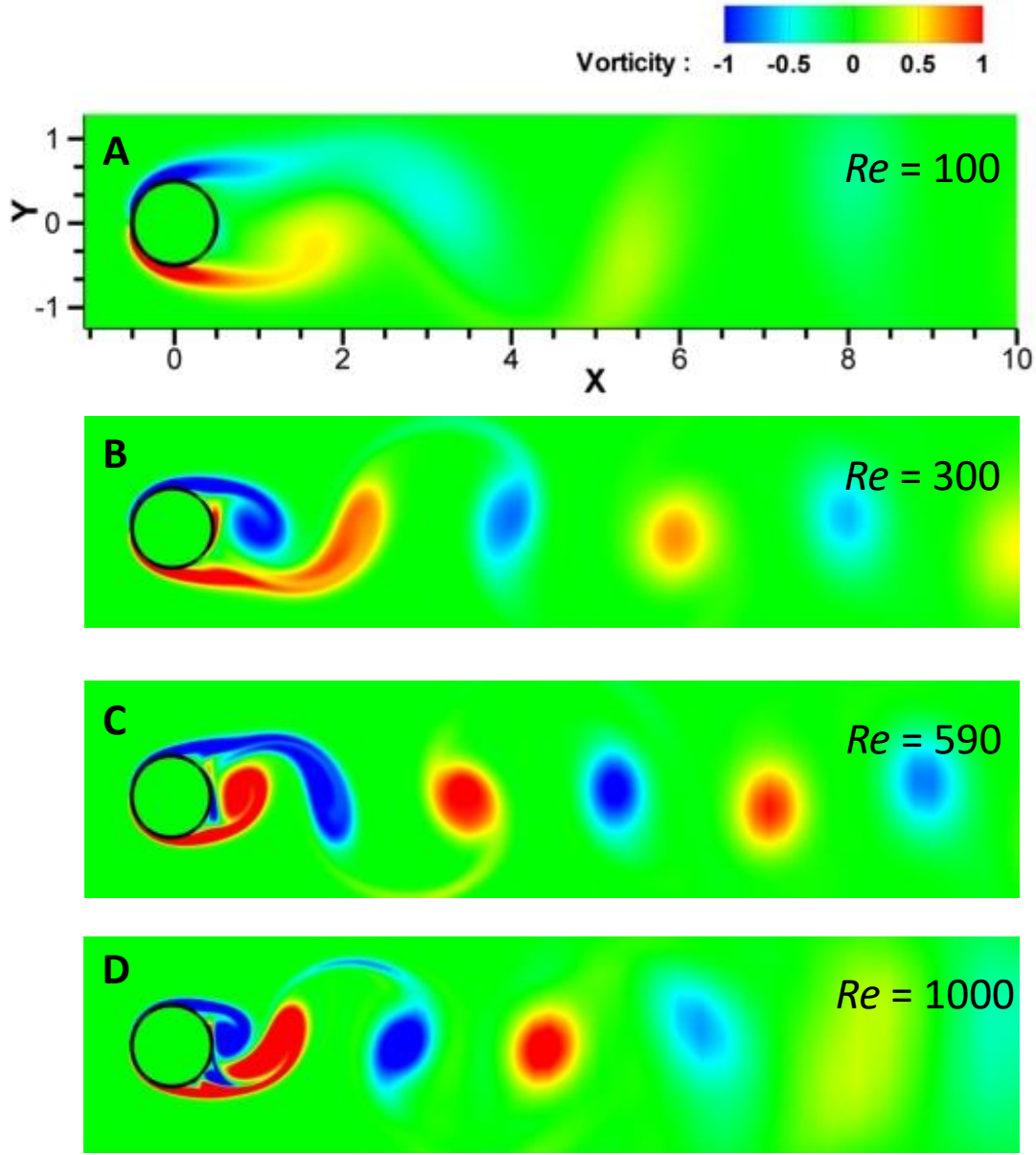

Fig. 7: Instantaneous vorticity contours at different Reynolds numbers (A) $Re = 100$, (B) $Re = 300$, (C) $Re = 590$ (D) $Re = 1000$.

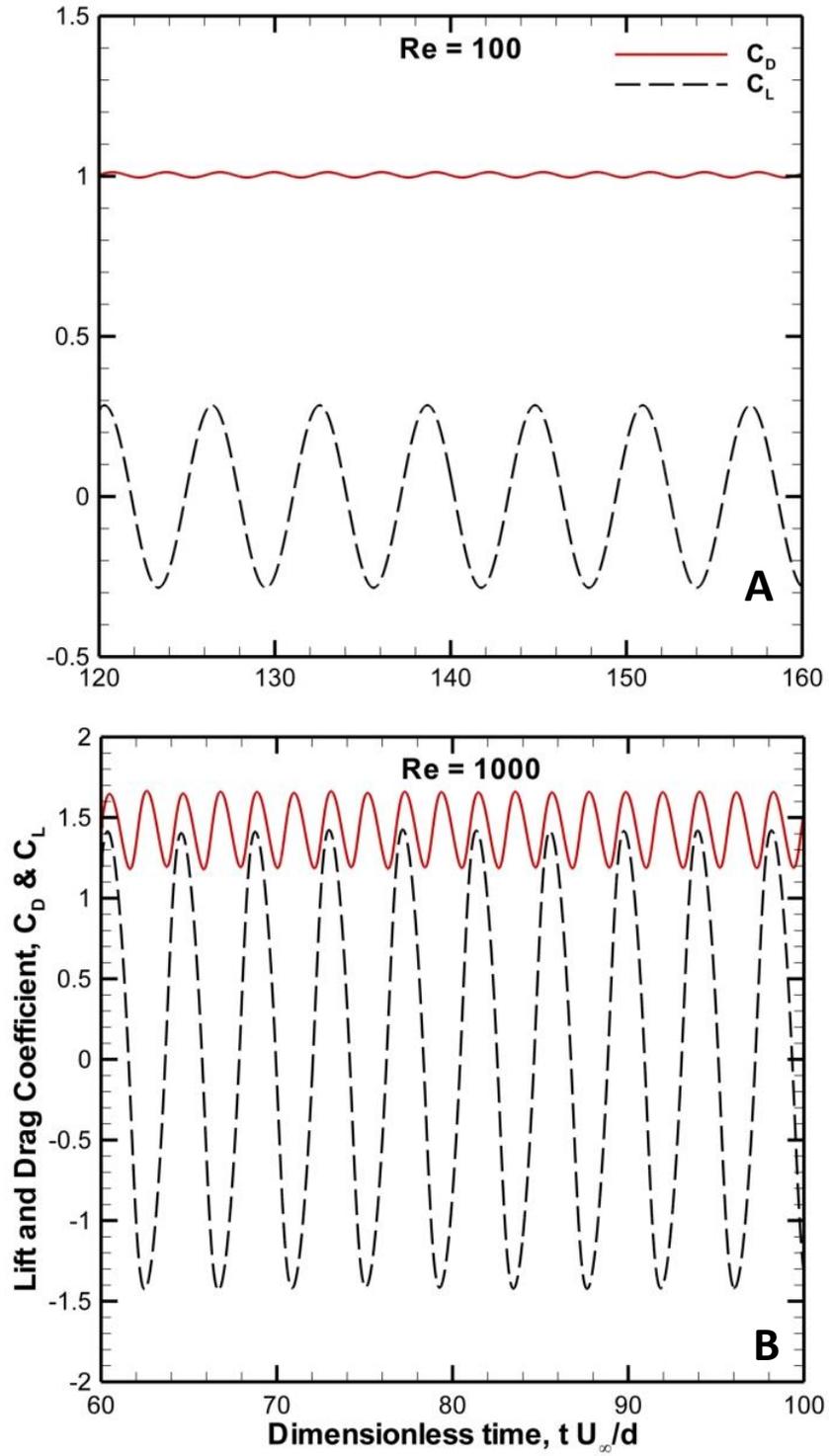

Fig. 8: Temporal variation of drag and lift coefficients at Reynolds number $Re = 100$ (A) and $Re = 1000$ (B).

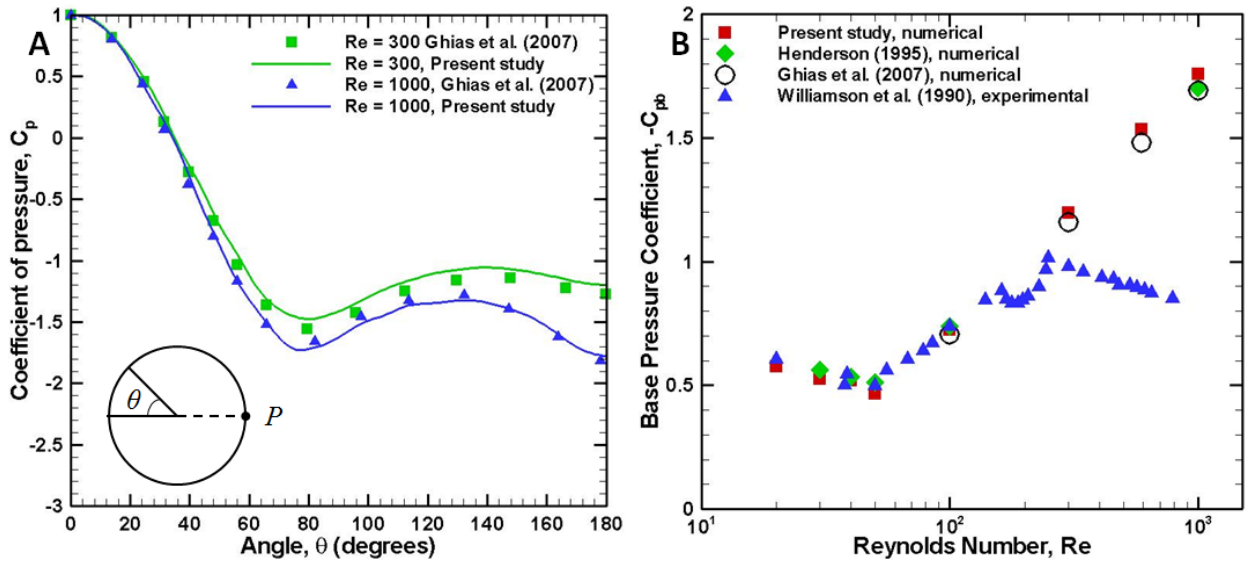

Fig. 9: (A) Comparison of coefficient of pressure on the cylinder surface at different Reynolds numbers with data of Ghias et al. [22] (B) Comparison of base pressure coefficient with the data of Ghias et al. [22] , Williamson et al. [41] and Henderson [40].

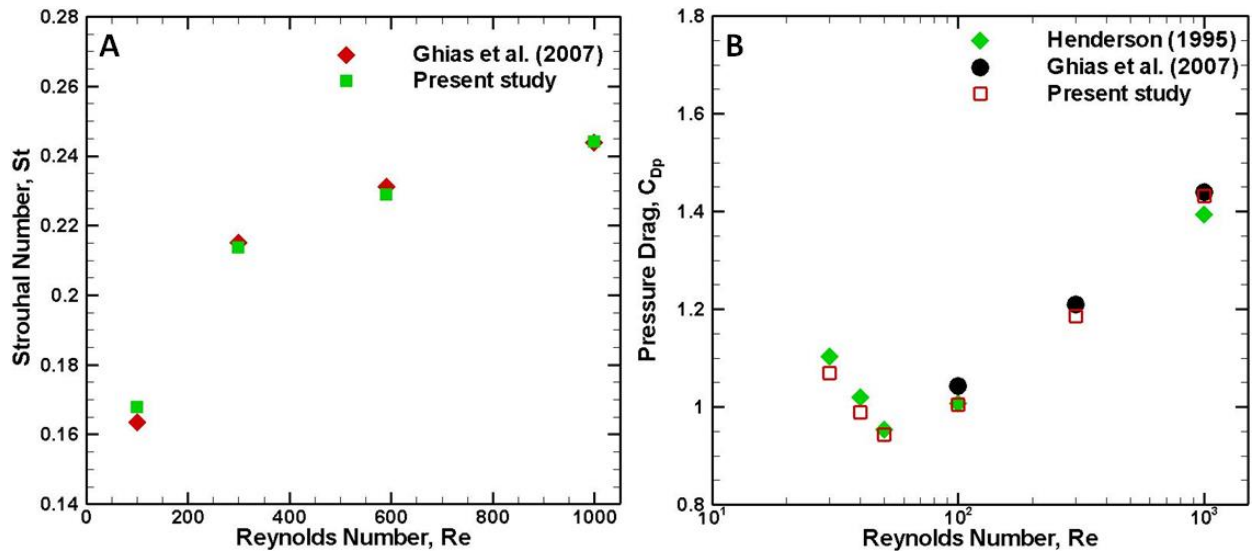

Fig. 10: (A) Comparison of Strouhal number at different Reynolds numbers with respective data of Ghias et al. [22] (B) Comparison of pressure drag coefficient with data of Ghias et al. [22] and Henderson [40].

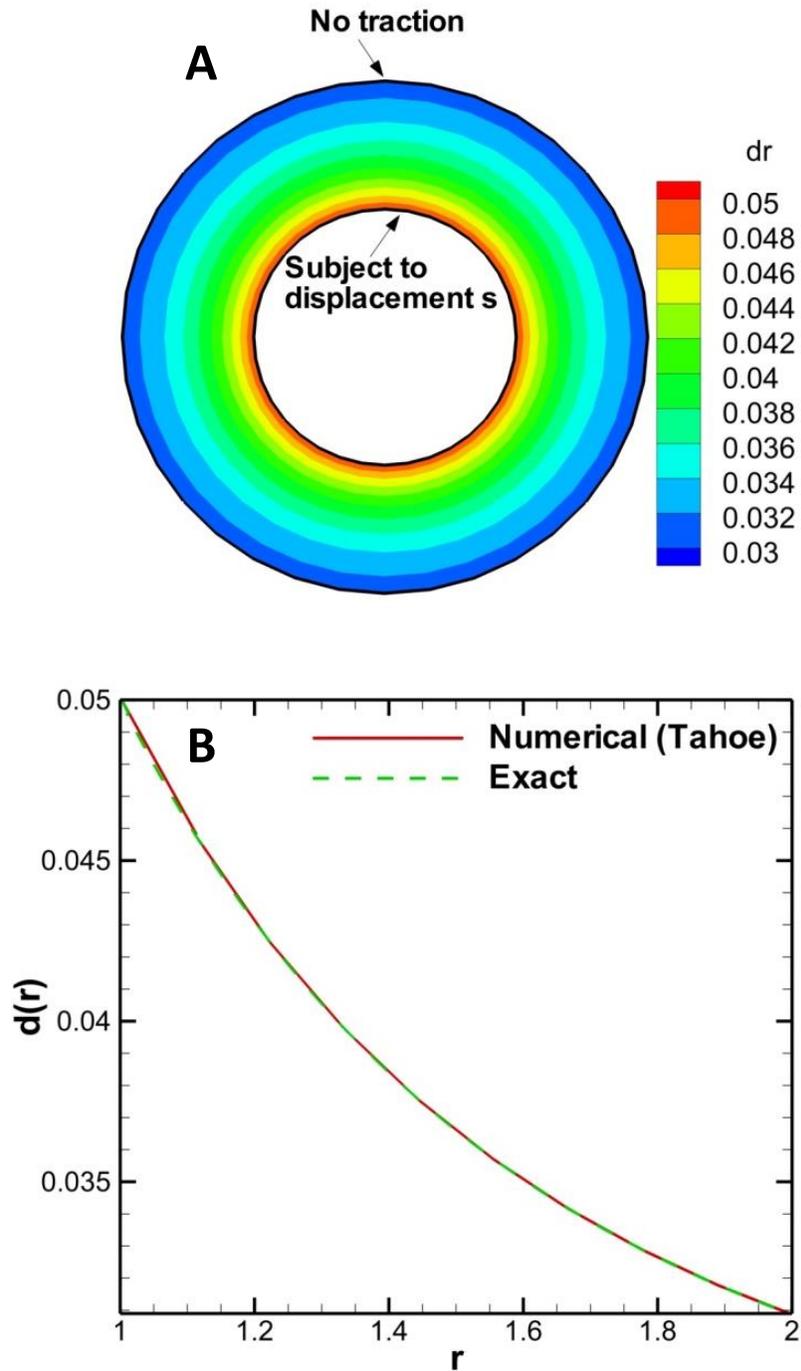

Fig. 11: Validation of structural solver for an infinitely long annulus subjected to prescribed radial displacement and zero traction at inner and outer surface, respectively. (A) Computed displacement contours by the structural solver (B) Comparison between the numerical and exact solution for the displacement profile along the thickness of the annulus.

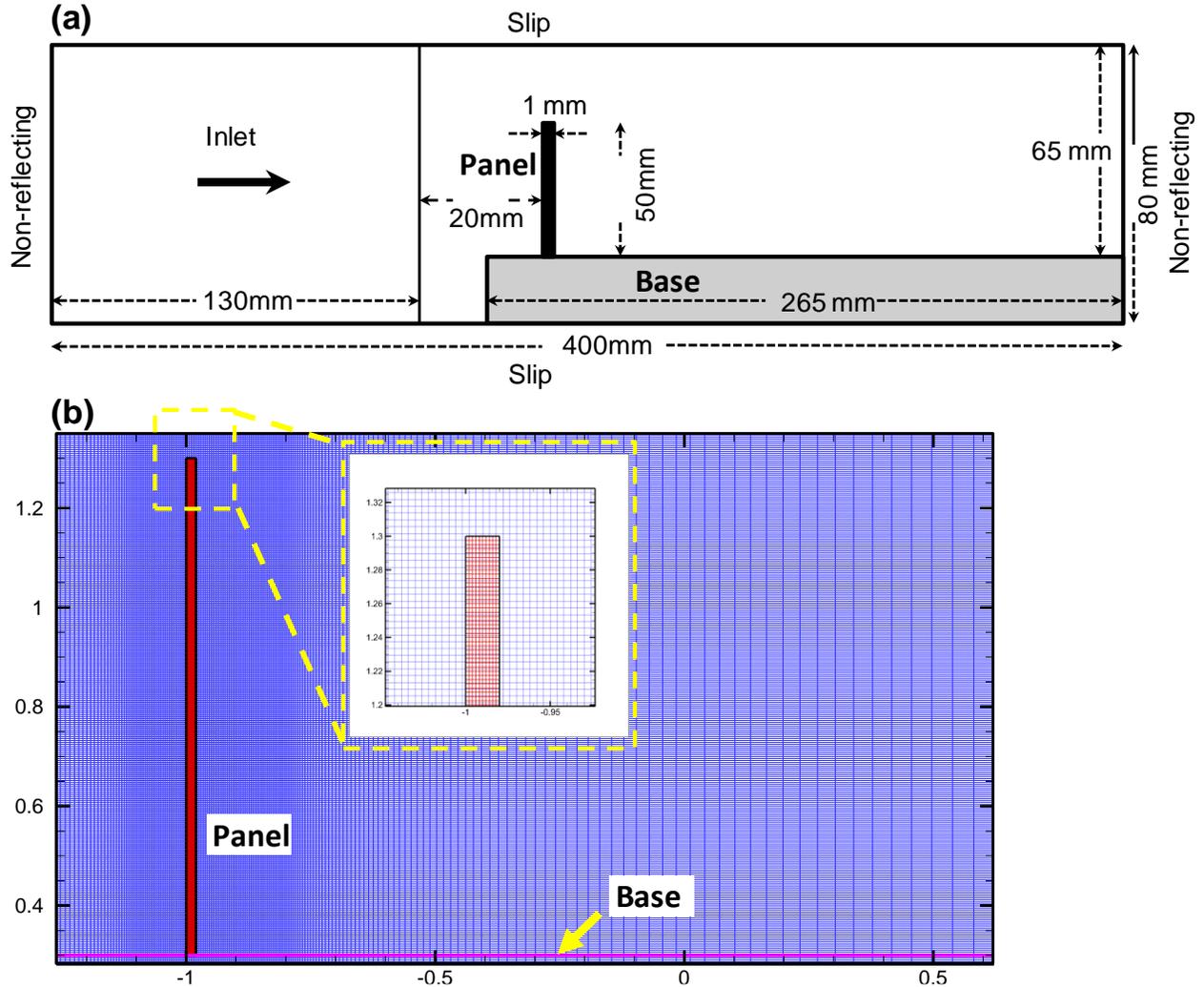

Fig. 12: (a) Computational domain used in the simulation setup of FSI benchmark problem proposed by Giordano et al. [43]. (b) Non-uniform Cartesian mesh with grid stretching in downstream employed in the present study. Finite-element mesh used in the panel and Cartesian mesh surrounding it are shown in the inset.

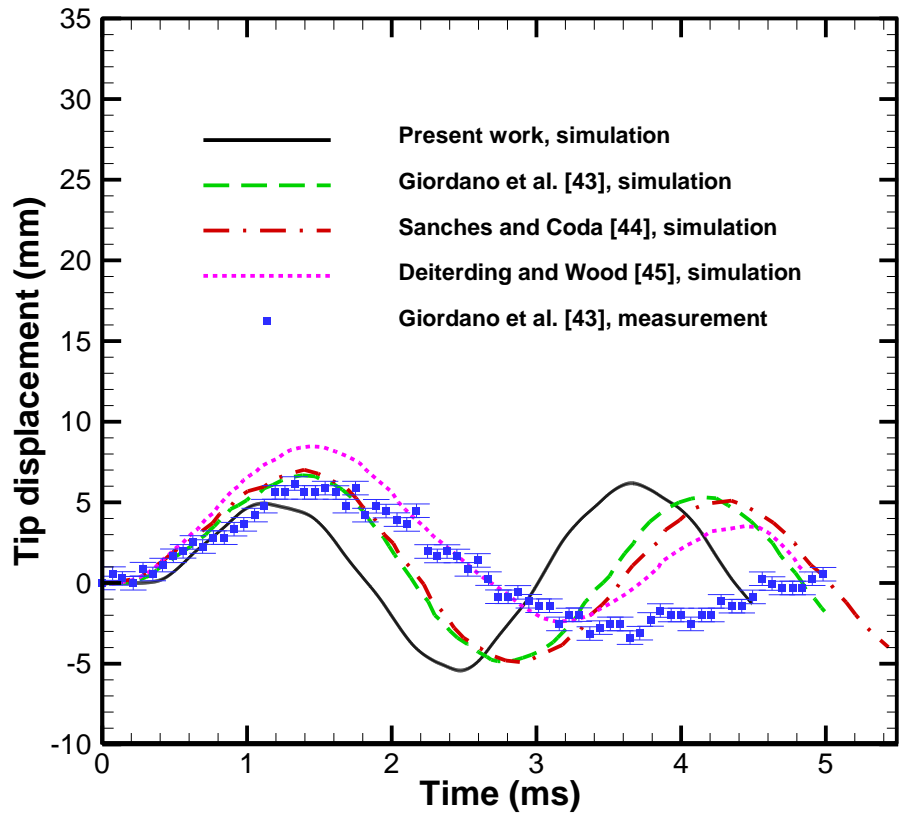

Fig. 13: Time-varying displacement of the tip of the panel obtained in the present simulation and its comparison with published experimental and numerical data.

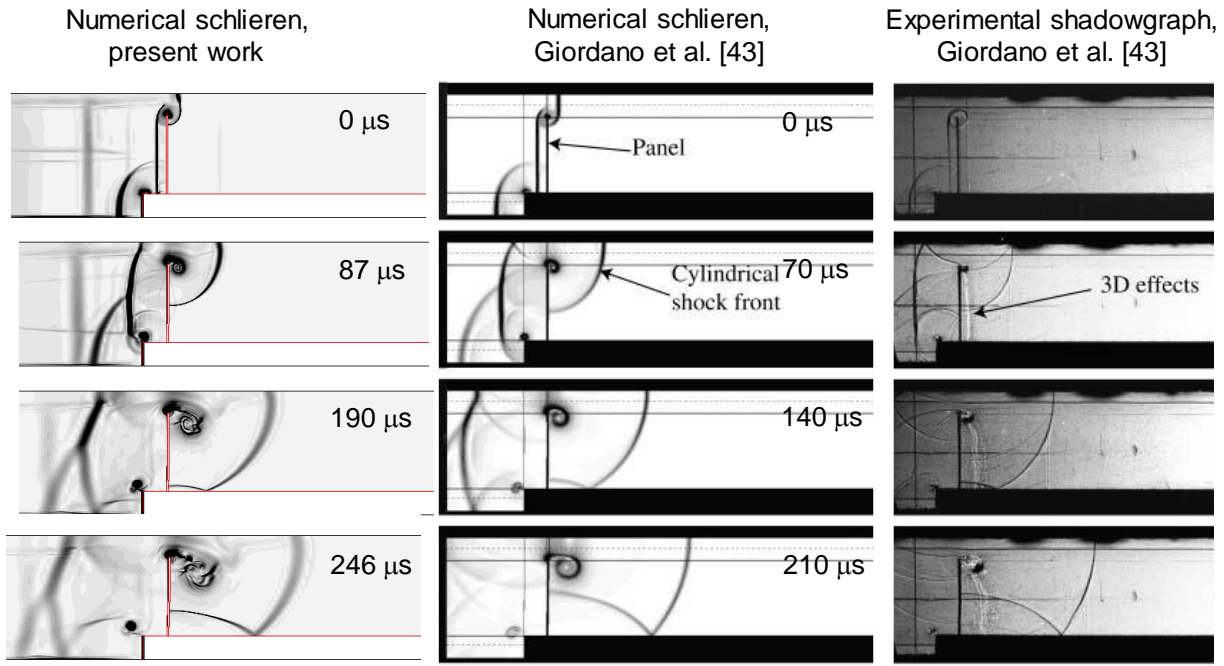

Fig. 14: (a) Qualitative comparison of computed shock propagation during the shock impact on elastic panel. First column shows snapshots of the contours of the computed magnitude of the density gradient (numerical schlieren) in the present work, while second and third column represent numerical schlieren and experimental shadowgraph, both reported by Giordano et al. [43]. Second and third column in Ref. [43] are adapted with permission (copyright Springer, 2005).

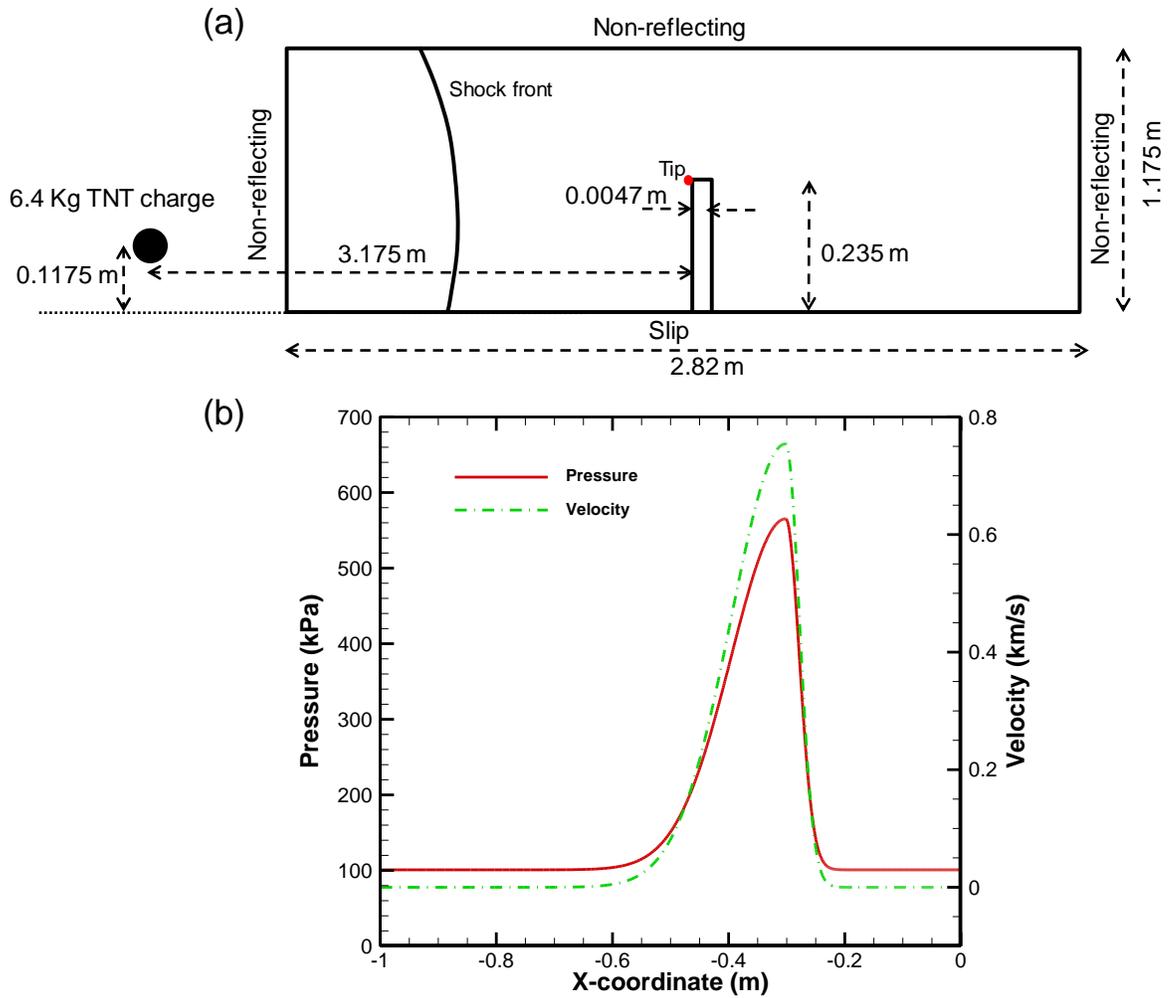

Fig. 15: (a) Computational domain, boundary conditions and initial position of the shock front in simulation setup of a realistic blast impact on a thin elastic plate. (b) Initial conditions of the shock front used in the computational domain.

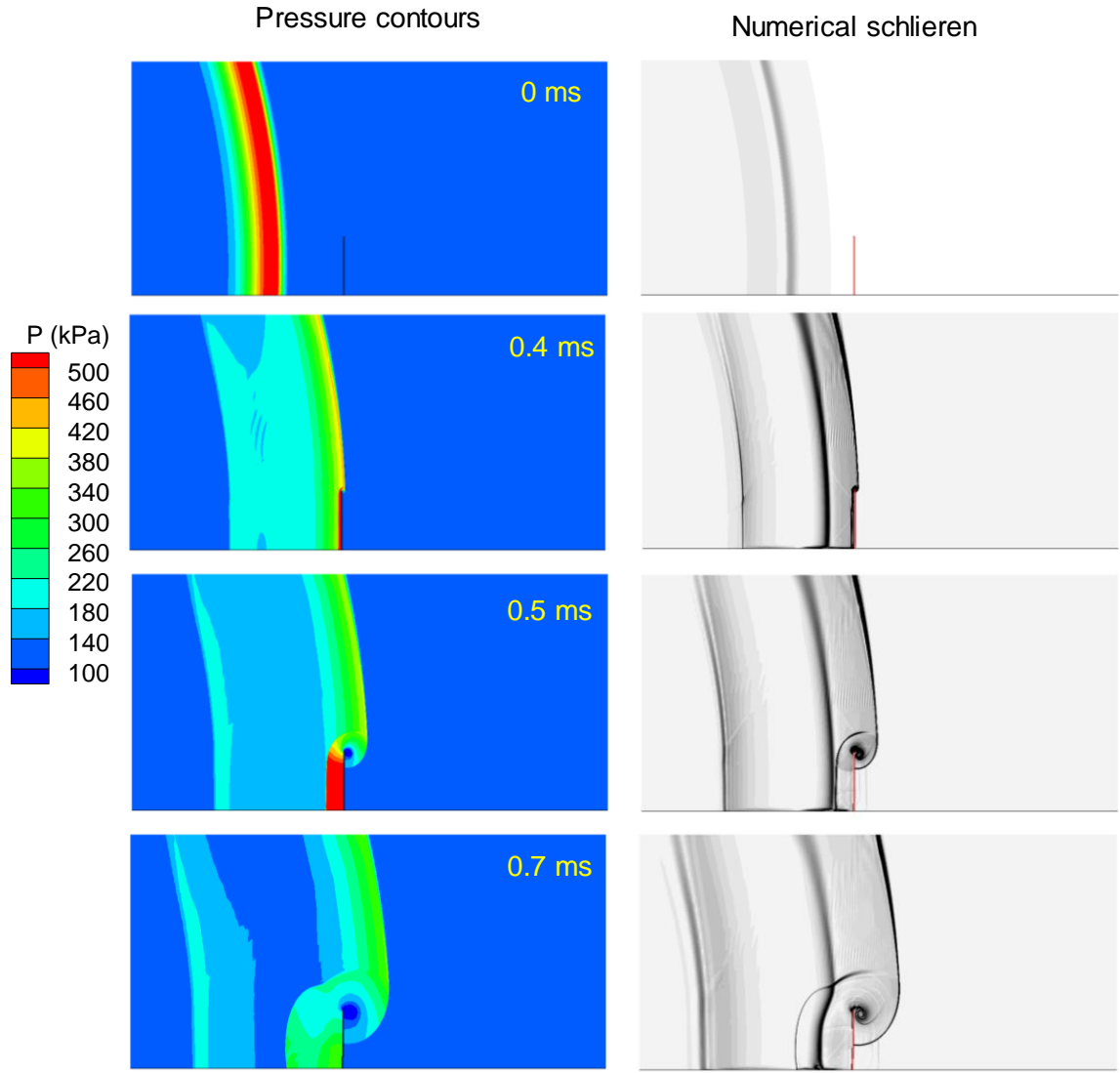

Fig. 16: Simulated pressure contours (left column) and numerical schlieren (right column) for 6.4 kg TNT blast on an aluminum plate at different time instances. Blast wave propagation, its interaction with the plate and Mach reflection can be noted.

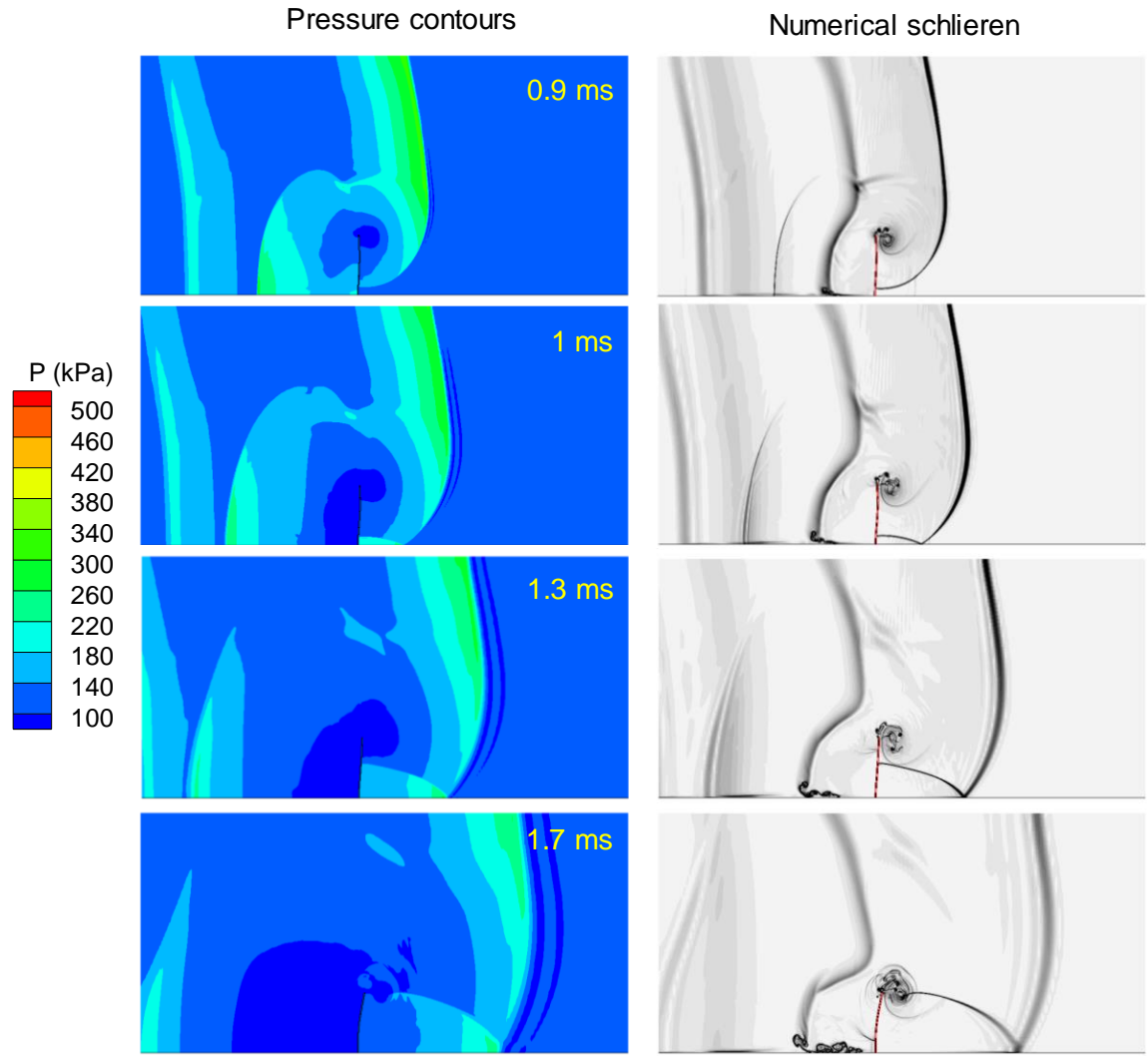

Fig. 17: Continuation of Fig. 16 for later times.

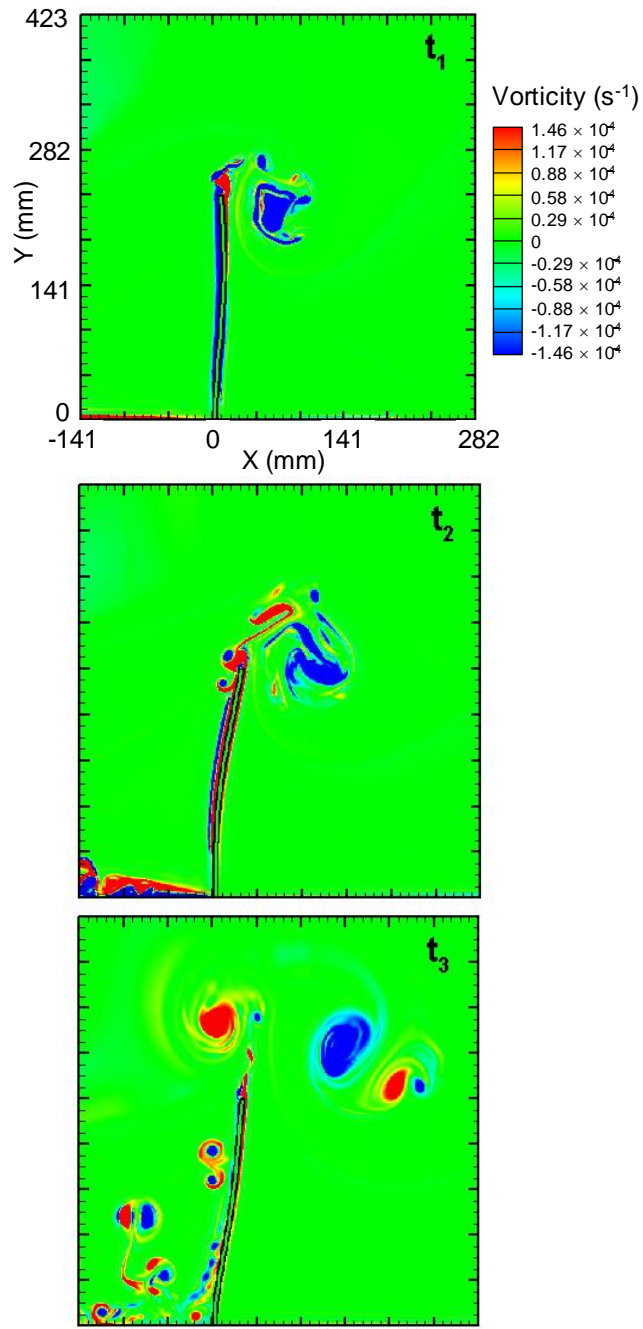

Fig. 18: Vorticity for the three time instances around the deforming plate. The three time instances are $t_1 = 1$ ms, $t_2 = 1.7$ ms and $t_3 = 3.1$ ms.

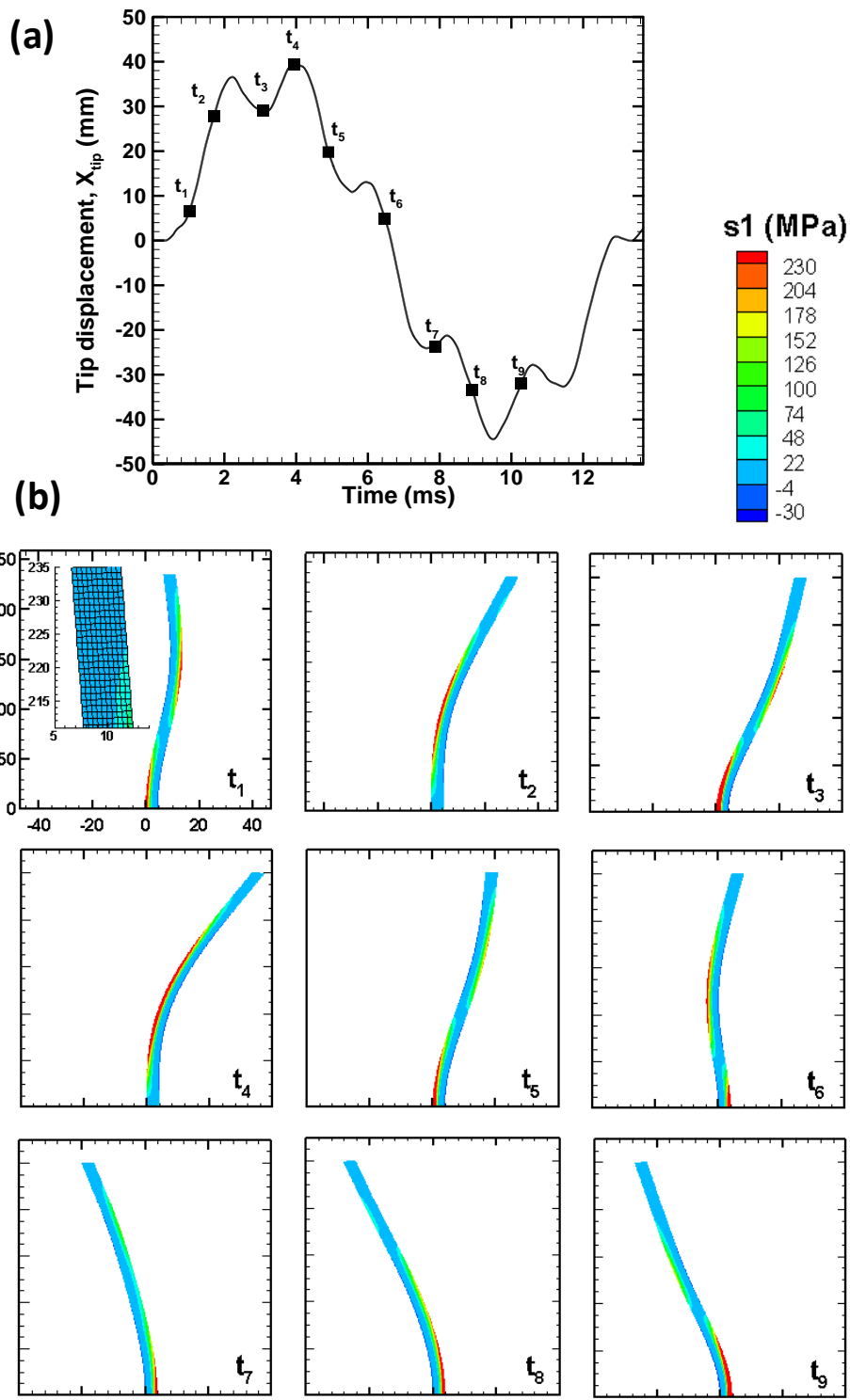

Fig. 19: (a) Time-varying displacement of the tip of the plate ($X_{\text{tip}}$) during the blast loading on the aluminum plate (b) Principal stress contours at different time instances in the aluminum plate. Inset at time $t_1$ shows finite-element mesh in the plate.

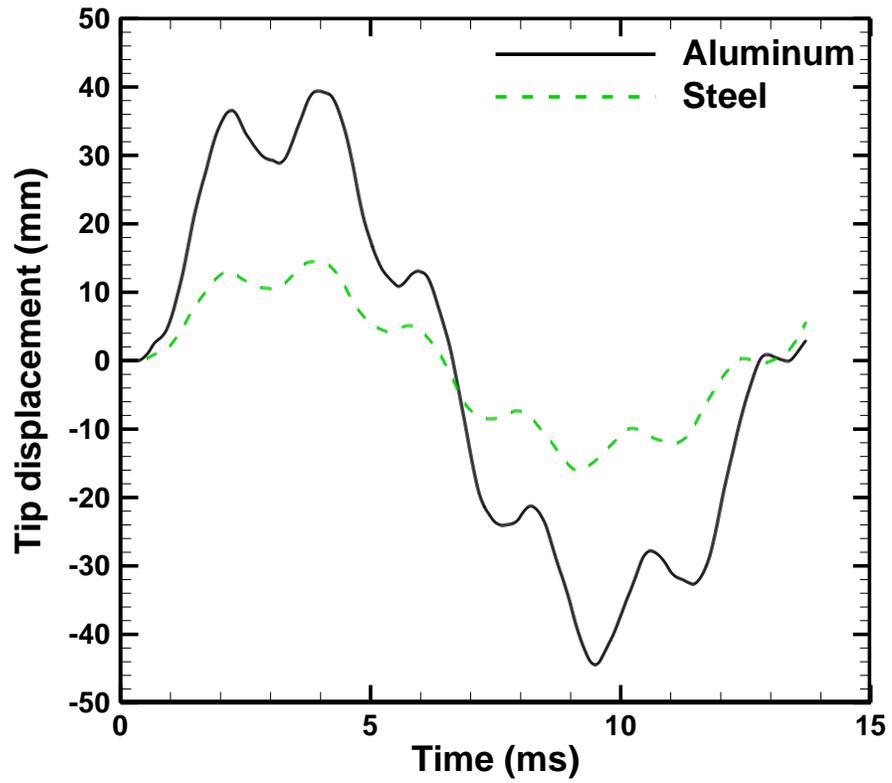

Fig. 20: Comparison between time-varying dimensionless tip displacement of the plates of aluminum and steel.